\documentclass[
superscriptaddress,
showkeys,
 amsmath,amssymb,
 aps,
prl,
twocolumn,
]{revtex4-1}

\usepackage{graphicx}
\usepackage{dcolumn}
\usepackage{bm}
\usepackage{epstopdf}
\usepackage[ansinew]{inputenc}
\usepackage{color}

\newcommand{\ew}[1]{\left\langle #1\right\rangle}
\newcommand{\ndg}{{\phantom{\dagger}}}
\newcommand{\dg}{\dagger}
\newcommand{\ket}[1]{\left|#1\right\rangle}
\newcommand{\bra}[1]{\left\langle #1\right|}

\begin{document}

\preprint{HOM with Variable Pulse Separation}

\title{Exploring Dephasing of a Solid-State Quantum Emitter via Time- and\\
 Temperature- Dependent Hong-Ou-Mandel Experiments}

\author{A.~Thoma}%
\affiliation{Institut für Festkörperphysik, Technische Universität Berlin,
Hardenbergstraße 36, 10623 Berlin, Germany}%
\author{P.~Schnauber}%
\affiliation{Institut für Festkörperphysik, Technische Universität Berlin,
Hardenbergstraße 36, 10623 Berlin, Germany}%
\author{M.~Gschrey}%
\affiliation{Institut für Festkörperphysik, Technische Universität Berlin,
Hardenbergstraße 36, 10623 Berlin, Germany}%
\author{M.~Seifried}%
\affiliation{Institut für Festkörperphysik, Technische Universität Berlin,
Hardenbergstraße 36, 10623 Berlin, Germany}%
\author{J.~Wolters}%
\affiliation{Institut für Festkörperphysik, Technische Universität Berlin,
Hardenbergstraße 36, 10623 Berlin, Germany}%
\author{J.-H.~Schulze}%
\affiliation{Institut für Festkörperphysik, Technische Universität Berlin,
Hardenbergstraße 36, 10623 Berlin, Germany}%
\author{A.~Strittmatter}%
\affiliation{Institut für Festkörperphysik, Technische Universität Berlin,
Hardenbergstraße 36, 10623 Berlin, Germany}%
\author{S.~Rodt}%
\affiliation{Institut für Festkörperphysik, Technische Universität Berlin,
Hardenbergstraße 36, 10623 Berlin, Germany}%
\author{A.~Carmele}%
\affiliation{Institut für Theoretische Physik, Technische Universität Berlin,
Hardenbergstraße 36, 10623 Berlin, Germany}%
\author{A.~Knorr}%
\affiliation{Institut für Theoretische Physik, Technische Universität Berlin,
Hardenbergstraße 36, 10623 Berlin, Germany}%
\author{T.~Heindel}%
\email{tobias.heindel@tu-berlin.de}
\affiliation{Institut für Festkörperphysik, Technische Universität Berlin,
Hardenbergstraße 36, 10623 Berlin, Germany}%
\author{S.~Reitzenstein}%
\affiliation{Institut für Festkörperphysik, Technische Universität Berlin,
Hardenbergstraße 36, 10623 Berlin, Germany}%

\date{\today}

\begin{abstract}
We probe the indistinguishability of photons emitted by a semiconductor quantum dot (QD) via time- and temperature- dependent two-photon interference (TPI) experiments. An increase in temporal-separation between consecutive photon emission events, reveals a decrease in TPI visibility on a nanosecond timescale, theoretically described by a non-Markovian noise process in agreement with fluctuating charge-traps in the QD's vicinity. Phonon-induced pure dephasing results in a decrease in TPI visibility from $(96\pm4)\,$\% at 10\,K to a vanishing visibility at 40\,K. In contrast to Michelson-type measurements, our experiments provide direct access to the time-dependent coherence of a quantum emitter at a nanosecond timescale.
\end{abstract}


\maketitle
Bright non-classical light sources emitting single indistinguishable photons on
demand constitute key building blocks towards the realization of advanced quantum
communication networks \cite{Knill2001,Kiraz2004,Kok2007,Gisin2007,Kimble2008}. In recent years, single self-assembled quantum dots (QDs) integrated into photonic
microstructures turned out to be very promising candidates for realizing
such quantum-light sources \cite{Michler2000,Santori2002b,Patel2008,Ates2009}, and enabled, for instance, a record-high photon indistinguishability of 99.5\,\% using self-organized InAs QDs under strict-resonant excitation \cite{Wei2014}.
Further advancement of quantum optical experiments and applications of QDs beyond proof-of-principle demonstrations, however, will certainly rely on deterministic device
technologies and should be compatible with scalable fabrication platforms. Furthermore, profound knowledge of the two-photon interference (TPI) is crucial for an optimization of novel concepts and devices in the field of advanced quantum information technology. In this respect, previous experiments utilizing QDs showed that dephasing
crucially influences the indistinguishability of the photons emitted by the QD
states, while a detailed understanding of the involved processes has been elusive \cite{Santori2004,Varoutsis2005,Gazzano2013,Gold2014}. In fact, these experiments revealed the difficulty of giving an adequate measure of the coherence time
$T_2$ of QDs. They even triggered a debate of how to correctly interpret $T_2$ obtained via Michelson interferometry, which typically gives a lower bound for the visibilities observed experimentally in Hong-Ou-Mandel (HOM) -type (\cite{Hong1987}) TPI experiments \cite{Santori2004,Gold2014,Bennett2005c}. A commonly accepted - although not proven - explanation for this apparent discrepancy is the presence of spectral diffusion on a timescale which is long compared to the excitation pulse-separation of a few nanoseconds typically used in HOM studies, but much shorter than the integration times of Michelson experiments. In this context, a more direct experimental access to the time dependent dephasing processes and their theoretical description is highly beneficial \cite{Wolters2013,Stanley2014}.

In this work, we map the coherence of a solid-state quantum emitter in the presence of pure dephasing by means of HOM-type TPI experiments. The timescale of the involved decoherence processes is precisely probed using an excitation sequence at which the temporal pulse-separation $\delta t$ is varried. 
Additionally, temperature-dependent measurements allow us to independently probe the impact of phonon-induced pure dephasing on the indistinguishabilty of photons.

\begin{figure}[t]
\includegraphics[width=1\linewidth]{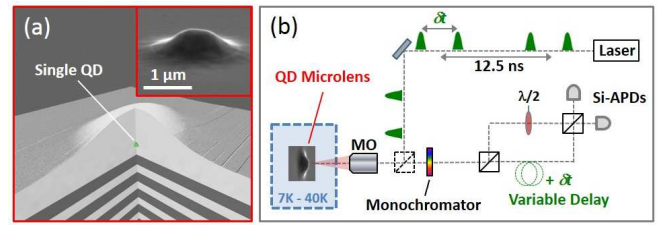}
\caption{\label{fig:fig_1} (a) Schematic view of the cross-section of a monolithic microlens with a single
deterministically integrated QD.
Inset: Scanning electron microscopy image of a fully processed microlens. (b)
Experimental setup: Hong-Ou-Mandel-type two-photon interference experiments are
utilized to probe the indistinguishability of consecutively emitted photons with variable pulse-separation $\delta t$.}
\end{figure}

The quantum emitter studied in our experiments is a single InAs QD grown by metal-organic chemical
vapor deposition (MOCVD) which is deterministically integrated within a monolithic microlens \cite{Gschrey2013,Gschrey2015a} (cf. Fig.~\ref{fig:fig_1}, see also Supplemental Material for details).
The quantum optical properties of photons emitted by the deterministic QD microlens are studied via low-temperature micro-photoluminescence spectroscopy in combination with HOM-type TPI experiments (cf. Fig.~\ref{fig:fig_1}\,(b), see Supplemental Material for experimental details). A mode-locked Ti:Sapphire laser operating in picosecond mode is used to excite the QD at a repetition rate of 80\,MHz. The periodic excitation pulses are converted to a sequence of double-pulses with variable pulse-separation $\delta t$. This excitation scheme in combination with a HOM-type asymmetric Mach-Zehnder interferometer enables us to probe the TPI visibility of two photons emitted by the QD as a function of the time elapsed between consecutive emission events.

\begin{figure}[t]
\includegraphics[width=1\linewidth]{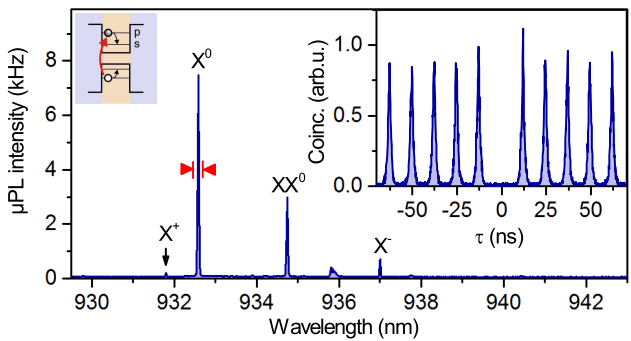}
\caption{\label{fig:fig_2} $\mu$PL spectrum of a deterministic QD microlens
under p-shell excitation ($T=7\,$K).
Inset: Second-order photon-autocorrelation measurement on the X$^0$ emission, demonstrating close to ideal single-photon emission.}
\end{figure}

A typical micro-photoluminescence ($\mu$PL) spectrum of a deterministic QD microlens chosen for our experiments is depicted in Fig.~\ref{fig:fig_2},
where the horizontally linearly polarized emission was selected using polarization optics.
The QD is excited pulsed ($\delta t=12.5\,$ns) quasi-resonantly in its p-shell at a wavelength of 909\,nm.
The assignment of the charge neutral exciton (X$^0$) and biexciton (XX$^0$) states as well as the charged trion states (X$^+$, X$^-$), was carried out via polarization and power dependent measurements as described e.g. in Ref. \cite{Schlehahn2015}.
For further investigations we first spectrally selected the emission of the X$^0$ state (cf. markers in Fig.~\ref{fig:fig_2}). The inset of Fig.~\ref{fig:fig_2} shows the corresponding raw measurement data of the second-order photon-autocorrelation $g^{(2)}_{\rm{HBT}}(\tau)$. 

In contrast to $g^{(2)}_{\rm{HBT}}(0)$, the photon-indistinguishability, being the crucial parameter for advanced quantum communication scenarios, is particular sensitive to dephasing processes. The dephasing rate of a quantum emitter is described by its coherence time $T_2$ and the radiative lifetime $T_1=\Gamma^{-1}$ via $T_2^{-1}=(2T_1)^{-1}+(T_2^*)^{-1}$ \cite{Bylander2003}, where $(T^*_2)^{-1}=\Gamma^\prime+\gamma$ describes pure dephasing due to spectral diffusion ($\Gamma^\prime$) and phonon interaction ($\gamma$).
In the following we gain experimental access to both
types of pure dephasing independently
by means of time- and temperature dependent TPI experiments. 

\begin{figure}[t]
\includegraphics[width=1\linewidth]{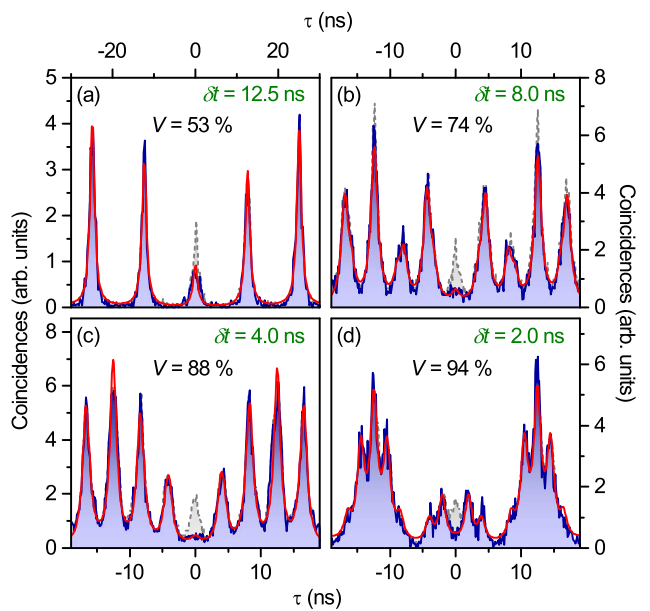}
\caption{\label{fig:fig_3} (a) to (d) Two-photon interference histograms
measured using a two-pulse excitation sequence with variable pulse-separation $\delta t$ ($T=7\,$K). Data
corresponding to co- (cross-) polarized measurement configuration are displayed by solid blue (dashed grey) curve, together with a fit to the data (solid red line) explained in the maintext.}
\end{figure}

First, we use a pulse sequence with 12.5\,ns
pulse-separation. Fig.~\ref{fig:fig_3}\,(a)
displays the obtained coincidence histogram of the two-photon detection events
at the two outputs of the HOM setup. In case of co-polarized photons (solid blue
curve), quantum-mechanical TPI manifests in a strongly reduced number of
coincidences at $\tau=0$, if compared to the measurement in cross-polarized
configuration (dashed grey curve). To quantitatively extract the visibility of TPI, we fitted Lorentzian profiles to the
experimental data in co-polarized configuration and evaluated the relative peak
areas according to Ref.~\cite{Santori2002b} (cf. Supplemental Material).
Under these excitation conditions, we extract a moderate visibility of $V_{\textup{12.5\,ns}}=(53\pm8)\,\%$.
A possible explanation for the finite wave packet overlap is an inhomogeneous spectral broadening of the QD transition due to spectral diffusion, leading to a pure dephasing rate
$\Gamma^\prime$ as mentioned above. Such processes are typically characterized by a certain timescale depending on specific material properties and growth
conditions
\cite{Empedocles1997,Robinson2000,Tuerck2000,Santori2002b,Bennett2005c,
Sallen2010,Houel2012}.

To perform a time-dependent analysis of $\Gamma^\prime$ and the underlying dephasing mechanism, we gradually reduce the pulse-separation $\delta t$ (vcf. Fig.~\ref{fig:fig_1}\,(b)), while the respective delay inside the HOM-interferometer is precisely matched to assure proper interference of consecutively emitted single photons. The resulting coincidence histograms for pulse-separations $\delta t$ of 8.0, 4.0 and 2.0\,ns are presented in
Fig.~\ref{fig:fig_3}\,(b) to (d).
The complex coincidence-pulse-pattern specific to each $\delta t$ results from overlapping
five-peak structures repeating every 12.5\,ns \cite{Mueller2014} (see Supplemental Material for details).
Fig.~\ref{fig:fig_4}\,(a)
\begin{figure}[t]
\includegraphics[width=1\linewidth]{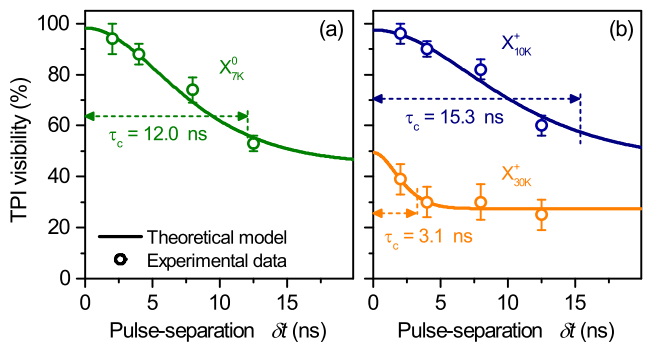}
\caption{\label{fig:fig_4} Two-photon interference visibilities of consecutively emitted single photons versus the time $\delta t$ elapsed between the emission
processes. Experimental data for (a) the X$^0$- and (b) the X$^+$-state are quantitatively described by a theoretical model assuming a non-Markovian noise correlation leading to spectral diffusion at a ns-timescale (see Eq.~\ref{eq:FinalResult}). A characteristic temperature-dependent correlation-time $\tau_c$ is observed.}
\end{figure}
summarizes the obtained raw TPI visibilities as a function of the
pulse-separation $\delta t$ for the neutral exciton X$^0$. At low $\delta t$ a plateau-like behavior is
observed, at which the visibility remains almost constant with values of
$V_{\textup{2.0ns}}=(94\pm6)\,\%$ and $V_{\textup{4.0ns}}=(88\pm4)\,\%$.
For pulse-separations larger than 4\,ns, a distinct
decrease in visibility is observed from $V_{\textup{8.0ns}}=(74\pm5)\,\%$ to $V_{\textup{12.5ns}}=(53\pm3)\,\%$.
The significant decrease in TPI visibility at pulse-separations larger than 8.0\,ns indicates the timescale of spectral diffusion. The time-dependent analysis of $\Gamma^\prime$ has additionally been carried out for the charged exciton state X$^+$ of the same QD at 10\,K and 30\,K (cf. Fig.~\ref{fig:fig_4}\,(b)). We observe again a characteristic correlation time, which decreases at higher temperature.

In order to gain deeper insight in the underlying dephasing mechanisms, we model the system with a Hamiltonian (see Supplemental Material),
where we approximated the QD as a two-level system with transition energy $\omega_e$.
To include dephasing, we employ the working horse of the phenomenological 
dephasing description by including a general stochastic force
$F(t) = P(t) + D(t)$ with a phonon-induced dephasing
($\delta$-correlated white noise) $P(t)$ and a spectral diffusion $D(t)$
component (colored noise), both shifting the transition energy of the QD.
The specific noise correlations depend on the coupling mechanism between the QD and its environment. For example, in case of spectral diffusion random electric fields due to charge fluctuations induce dephasing \cite{Tuerck2000,Galperin2006}, as discussed later on.
Given that the classical (pump) field excites the QD fast enough to prevent
multiple photon emission processes, we calculate via the
Wigner-Weisskopf method the wave function after the two pulse sequence:
\begin{align}
\notag
\ket{\Psi(t)} = \int_0^t dt_1 \int_{\delta t}^t dt_2
& e^{i(\omega_e+i\Gamma)(t_1+t_2)-i\phi_{\delta t}(t_2)-i\phi_0(t_1)} \\
& \times E_2(t_2) E_1(t_1) \ket{\text{vac}} \ .
\end{align}
This wave function includes the two-photon wave packages $E_n(t_n)$ and the
time-integrated stochastic forces defined as $\phi^X_i(t):= \int_i^t \text{d}t^\prime X_i(t^\prime)$,
where $i=0$ in case the photon was emitted during the first sequence or
$i=\delta t$ for photon emission processes due to the second pulse and $X(t)$ denoting the noise.
Considering the interference at the beamsplitter by unitary transformations on the incident electric fields allows us to calculate the two-photon correlation $\ew{E^{(-)}_A(t) E^{(-)}_B(t+\tau) E^{(+)}_B(t+\tau) E^{(+)}_A(t)}$ measured in the experiment at detector A and B.
To evaluate the stochastic forces, we need to average via a Gaussian 
random number distribution $ \left\langle\left\langle \cdot\right\rangle\right\rangle $. The $ \left\langle\left\langle \cdot\right\rangle\right\rangle $ denotes statistical averaging in terms of a Gaussian random variable, where all higher moments can be expressed by the second-order correlation \cite{Gardiner2004}.
Here, we employ the simplest possible model described as a Markovian process
$\delta-$correlated in time, i.e. as white noise.
It is highly temperature dependent and limits the 
absolute value of the indistinguishability, independent from the temporal
distance of the excitation pulses $\delta t$.
In contrast to the phonon-induced dephasing, the spectral diffusion reveals
a strong dependence on the pulse distance, as seen Fig.~\ref{fig:fig_4}.
We include this dependence as a finite memory-effect with specific 
correlation time $\tau_c$: 
\begin{align}
\notag
\left\langle\left\langle  
\phi^D_{t_1}(t_2) \phi^D_{t_3}(t_4)
\right\rangle\right\rangle  
= 
\int_{t_1}^{t_2}
dt
\int_{t_3}^{t_4}
dt^\prime 
\left\langle D(t) D(t^\prime) \right\rangle \\
= \Gamma^\prime_0 \ e^{-\frac{(t_1-t_3)^2}{\tau_c^2}} \left( \text{min}[t_2,t_4] - \text{max}[t_1,t_3] \right) \ ,
\end{align}
where $\Gamma^\prime_0$ describes the maximal amount of pure dephasing induced by spectral diffusion.
These kinds of noise correlations stem from a non-Markovian low-frequency 
noise \cite{Galperin2006,Laikhtman1985,Eberly1984} and show plateau-like behavior for temporal pulse distances sufficiently
short in comparison to the memory depth. Thus, if $ \delta t \ll \tau_c $, the
effect of spectral diffusion becomes negligible and phonon-induced
dephasing limits the absolute value of the visibility.
Using these correlations, assuming a balanced beamsplitter ($R=T=1/2$) and normalizing the two-photon
correlation we derive the following formula, which explicitly depends on the pulse-separation $\delta t$:
\begin{align}
V(\delta t,\tau_c,T) =\frac{\Gamma}{\Gamma^\prime_0(1-e^{-(\delta t/\tau_c)^2})+\gamma(T) + \Gamma} \ .
\label{eq:FinalResult}
\end{align}
Here, $\Gamma^\prime:=\Gamma_0^\prime(1-e^{-(\delta t/\tau_c)^2})$ corresponds to the $\delta t$-dependent pure dephasing due to spectral diffusion.
In case of vanishing phonon-induced dephasing and spectral diffusion, the TPI
visibility is $1$, i.e. the photons are Fourier-transform-limited and
coalesce at the beamsplitter into a perfect coherent two-photon state.
For low temperatures, the phonon-induced dephasing is small and the spectral
diffusion with a finite memory depth dictates the functional form of the
visibility for different pulse distances.

Applying the model derived in Eq.~\ref{eq:FinalResult} to the experimental data of Fig.~\ref{fig:fig_4}, by fixing $\Gamma$ (measured independently via time-resolved measurements) and assuming $\gamma_{\textup{7K,10K}}=0$ (cf. next paragraph), we deduce correlation times $\tau_c$ listed in Tab.~\ref{tab:table1}. The timescale at which the noise is correlated appears to be close to the fundamental period of the Ti:Sapphire laser for X$^0_{\textup{7K}}$ and X$^+_{\textup{10K}}$,  whereas an increase in temperature to 30\,K shortens the correlation time of X$^+$ drastically (cf. Tab.~\ref{tab:table1}).
Interestingly, the coherence times $T_2^{\infty}$ inferred from our model in the limit $\delta t\rightarrow\infty$ (see Tab.~\ref{tab:table1}), significantly exceed the values of $T_{2}=(291\pm6)\,$ps for X$^0_{7K}$ and $T_{2}=(167\pm3)\,$ps for X$^+_{30K}$ obtained via measurements using a Michelson-interferometer (see Supplemental Material).
\begin{table}[t]
\caption{\label{tab:table1}%
Correlation times $\tau_c\,$ obtained by fitting Eq.~\ref{eq:FinalResult} to the experimental data of Fig.~\ref{fig:fig_4}, fixing $\gamma_{\textup{7K,10K}}=0$ and $\Gamma$. $T_2^{\infty}$ values have been calculated from the parameters $\Gamma$, $\Gamma^\prime_0$ and $\gamma$.}
\begin{ruledtabular}
\begin{tabular}{lccccc}
\textrm{}&
\textrm{$\Gamma\,$(GHz)}&
\textrm{$\Gamma^\prime_0\,$(GHz)}&
\textrm{$\gamma\,$(GHz)}&
\textrm{$\tau_c\,$(ns)}&
\textrm{$T_2^{\infty}\,$(ps)}\\
\colrule
\\
X$^0_{\textup{7K}}$ & 0.85 & $1.02\pm0.06$ & 0 & $12.0\pm1.9$ & 692 \\
X$^+_{\textup{10K}}$ & 0.91 & $1.03\pm0.04$ & 0 & $15.3\pm2.5$ & 673 \\
X$^+_{\textup{30K}}$ & 0.96 & $1.55\pm0.78$ & $0.29\pm_{0.29}^{1.1}$ & $3.1\pm1.9$ & 431 
\end{tabular}
\end{ruledtabular}
\end{table}
A physical origin of the plateau-like behavior of  $V(\delta t)$ and the associated non-Markovian decoherence processes are random flips
of bistable fluctuators in the vicinity of the QD \cite{Laikhtman1985}.
Possible candidates for such fluctuators in solid state devices are charge
traps or structural dynamic defects \cite{Galperin2006}. Further evidence for the presence of charge fluctuations is given by the observation of trion states X$^+$ and X$^-$ under quasi resonant excitation of the QD (cf. Fig.~\ref{fig:fig_2}). To reduce the associated electric field noise, weak optical excitation above-bandgap \cite{Gazzano2013} or a static electric field via gates \cite{Stanley2014} can be applied.

\begin{figure}[t]
\includegraphics[width=1\linewidth]{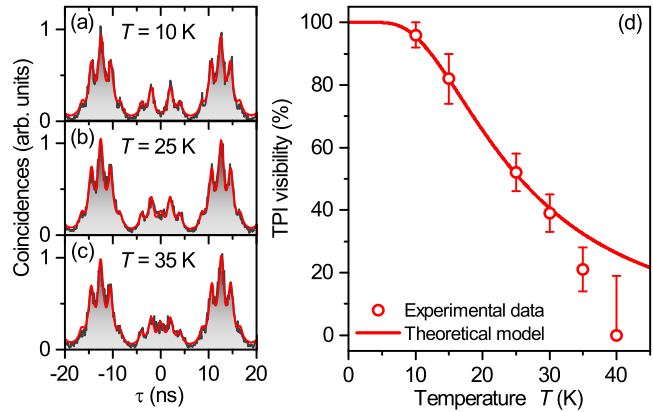}
\caption{\label{fig:fig_5} Impact of the temperature on the two-photon
interference (TPI) visibiliy ($\delta t=2\,$ns). (a)-(c) TPI histograms for co-polarized configuration at 10,
25 and 35\,K and corresponding fits (red solid curves). (d) Experimentally obtained TPI visibilities for various temperatures. We achieve qualitative agreement with a theoretical model assuming dephasing proportional to the phonon number (see Supplemental Material).}
\end{figure}

To justify the assumption $\gamma_{\textup{7K,10K}}=0$ and to investigate the influence of phonons on the photon-indistinguishability in more detail, we performed complementary temperature dependent TPI experiments.
For this purpose, the emission of the trion state X$^+$ was selected under quasi-resonant excitation and coupled to the HOM-interferometer. The pulse-separation was fixed to $\delta t=2.0\,$ns, while the temperature $T$ was varied. Fig.~\ref{fig:fig_5}\,(a) to (c)
exemplarily display TPI coincidence histograms for temperatures $T$
of 10, 25 and 35\,K in co-polarized measurement configuration. A
gradual increase in coincidences at $\tau=0$ is observed, indicating a reduced photon-indistinguishability. The obtained TPI visibilities extracted from the
experimental data for temperatures ranging from 10 to 40\,K are depicted in
Fig.~\ref{fig:fig_5}\,(d). At low temperature, we observe close to ideal photon-indistinguishability with $V_{\textup{10K}}=(96\pm4)\,$\%. Increasing $T$ results in a distinct decrease of the TPI visibility. Finally, at a temperature of $40\,$K, $V$ approaches zero within the standard error of our measurement. The observed temperature dependence is further modeled theoretically (red solid line).
For this purpose we employed a Markovian approximation for the phonon-induced pure dephasing processes, where the dephasing is proportional to the square of the phonon number \cite{Carmichael1999} (see Supplemental Material for details). The model qualitatively describes our experimental observation. Hence, we conclude that the impact of $\gamma$ in Eq.~\ref{eq:FinalResult} is indeed almost negligible at low temperatures ($T\leq10\,$K), but has severe impact at elevated temperatures. For temperatures above $30$\,K, also in- and outscattering with wetting layer carriers needs to be included, which explains the slight deviation between experiment and theory in this temperature range.

In summary, we presented a method to directly access the time-dependent coherence of a single quantum emitter via HOM-type TPI experiments.
We explored the photon-indistinguishability as a function of the time $\delta t$ elapsed between consecutive photon emission events and for different temperatures. We observe TPI visibilities close to unity ($V_{\textup{10K}}=(96\pm4)\,$\%) for MOCVD-grown QDs under p-shell excitation at $\delta t=2.0\,$ns. Increasing $\delta t$ results in a decrease in visibility on a nanosecond timescale.
Our theoretical analysis shows that such behavior can be explained by a non-Markovian dephasing process, which is attributed to spectral diffusion caused by fluctuating charge traps. We independently study the impact of phonon-induced pure dephasing on the photon-indistinguishability.
Our findings have important implications with respect to the quantum interference of photons emitted by remote emitters \cite{Patel2010,Flagg2010,Bernien2012,Gold2014} and single-photon multiplexing schemes \cite{Ma2011}. 

\begin{acknowledgments}
We gratefully acknowledge expert sample preparation by R. Schmidt, and thank C.
Schneider and C. Matthiesen for stimulating discussions. This work was
financially supported by the German Research Foundation (DFG) within the Collaborative
Research Center SFB\,787 'Semiconductor Nanophotonics: Materials, Models,
Devices' and the German Federal Ministry of Education and Research (BMBF)
through the VIP-project QSOURCE (Grant No. 03V0630). A.C. gratefully acknowledges
support from the SFB\,910: 'Control of self-organizing nonlinear systems'.
\end{acknowledgments}

\appendix

\section{Supplemental Material}

\begin{figure}[t!]
\includegraphics[width=1\linewidth]{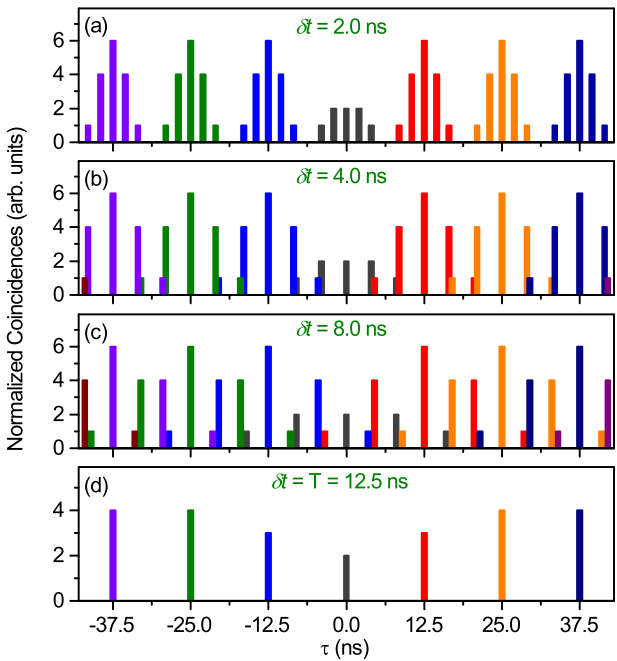}
\caption{\label{fig:fig_Appendix_1} Schematic coincidence pulse patterns
resulting from a two-pulse sequence with a pulse-separation $\delta t$ repeating
every $T=12.5\,$ns. The expected peak area ratios in case of distinguishable
photons are encoded in the the height of each bar.}
\end{figure}

\textbf{Sample growth and processing:}
The QD sample utilized for our experiments was grown by metal-organic chemical
vapor deposition (MOCVD) on GaAs (001) substrate. A low-density layer of
self-organized InGaAs QDs is deposited above a lower distributed Bragg reflector
(DBR) constituted of 23 alternating  $\lambda/4$-thick bi-layers of AlGaAs/GaAs.
On top of the QDs, a 400\,nm thick GaAs capping layer provides the material for
the subsequent microlens fabrication. To process monolithic single-QD
microlenses we used a recently developed deterministic technique exploiting
cathodoluminescence (CL) spectroscopy and 3D in-situ electron-beam lithography
\cite{Gschrey2013,Gschrey2015a}. Here, the sample is first spin-coated with a
190\,nm thick layer of polymethyl methacrylate (PMMA) acting as electron-beam
resist. Afterwards, CL intensity maps are recorded in a custom-build CL-system
at cryogenic temperature (5\,K) and low electron dose. Specific target QDs
are then selected for the integration into microlenses. For this purpose, lens-patterns are embossed into the resist by writing concentric
circles centered at the target QD's position, where the applied electron dose is
varied from highest values at the center to lowest values at the edge of the
microlens. Afterwards, the sample is transfered out of the CL-system, to develop the resist at room temperature. At this point the inverted (unsoluble) PMMA remains above target QDs and acts as a lens-shaped etch-mask, while the resist is completely removed in the remaining CL mapping region. Finally, the microlens
profile is transfered into the semiconductor material via dry etching using
inductively-coupled-plasma reactive-ion etching (ICP-RIE).
An SEM image of a readilly processed microlens is shown in Fig.~1\,(a) in the maintext. We have chosen shallow hemispheric microlens sections with heights of 400\,nm and base widths of 2.4\,$\mu$m, allowing for a photon extraction efficiency of 29\,\% \cite{Schlehahn2015a}.\\ 
\\
\textbf{Experimental setup:}
The experimental setup is based on $\mu$PL
spectroscopy in combination with HOM-type TPI experiments (cf.
Fig.~1\,(b) in main text). The QD-microlens chip is mounted onto the coldfinger
of a liquid-Helium-flow cryostat at cryogenic temperatures $T$ from 7 to
40\,K. A mode-locked Ti:Sapphire laser operating in picosecond mode with a
repetition rate of 80\,MHz is used to quasi-resonantly excite a single QD state
in its p-shell. The periodic optical pulses delivered by this laser system are
converted to a sequence of double-pulses with pulse-separation of $\delta t$ by
utilizing an asymmetric Mach-Zehnder interferometer based on polarization
maintaining (PM) single-mode fibers (not shown). By choosing different
fiber-delays within one arm of the interferometer, $\delta t$ can be varied
from 2.0\,ns up to 12.5\,ns. This two-pulse sequence is then launched onto a
single-QD microlens via a microscope objective (MO) with a numerical aperture of 0.4. 
The same MO is used to collect and collimate the QD's emission, which is subsequently focused onto the
entrance slit of an optical-grating monochromator with attached charge-coupled
device camera (spectral resolution: 0.017\,nm (25\,$\mu$eV)). Polarization
optics (linear polarizer and $\lambda/2$-waveplate) in front of the spectrometer
allow for polarization selection of particular QD states. To perform HOM-type
TPI experiments, a second PM-fiber-based asymmetric Mach-Zehnder interferometer
is attached to the output port of the spectrometer. Using a
$\lambda/2$-waveplate, the polarization of the photons in one interferometer arm
can be switched either being co- or cross-polarized with respect to the other
arm. To interfere consecutively emitted single photons at the second
beam-splitter, a variable fiber delay matched to the respective pulse-separation
$\delta t$ is implemented in one interferometer arm. The photon arrival time at
the second beamsplitter can be fine-tuned with a precision of 3\,ps. Finally, 
photons are detected at the two interferometer outputs using Silicon-based avalanche photodiodes (APDs) and
photon coincidences are recorded via time-correlated single-photon counting
(TCSPC) electronics enabling coincidence measurements with an overall timing
resolution of 350\,ps.\\
\\
\textbf{Evaluation of Visibility:}
To extract the TPI visibilities from the coincidence histograms obtained for
co-polarized measurement configuration (cf. Fig.~3 in maintext), the peak area ratios can be considered \cite{Mueller2014}.
Fig.~\ref{fig:fig_Appendix_1} 
schematically illustrates the coincidence pulse patterns resulting from the
applied two-pulse sequences with pulse-separations $\delta t$.
The peak area ratios deduced from the probability distribution of all possible pathway
combinations are represented by the respective bar height. Each pattern is
composed of five-peak clusters with temporal delays of $T=12.5\,$ns according to
the laser's fundamental repetition rate. The five-peak cluster in turn arises from the possible pathway-combinations taken by two
photons separated by $\delta t$. Thus, the peak area ratios can easily be deduced considering combinatorics, which enables us to extract the TPI visibility quantitatively.
The expected peak area ratio of each cluster is 1:4:6:4:1, except for the cluster
centered at zero-delay ($\tau=0$). Here, the peak area ratio depends on the photon-indistinguishability. In case of perfect
indistinguishability, the coincidences at $\tau=0$ vanish and the peak area
ratios of the cluster becomes 1:2:0:2:1.
Photons which are distinguishable, e.g. due to their polarization lead to an area ratio of 1:2:2:2:1. In the following, the peak areas of the central cluster are labeled
A$_2'$:A$_1'$:A$_0$:A$_1$:A$_2$ and $\bar{A}=(A_1'+A_1)/2$.
The corresponding peak areas are extracted from the measurement data by fitting
Lorentzian peaks with the expected area ratios to the coincidence histograms.
In all fits, we fixed the width of the Lorentzian peaks to the value obtained
from the fit to the data at $\delta t=T=12.5\,$ns. The TPI visibility for
$\delta t=2$, 4 and 8\,ns is then given by
\begin{eqnarray}
V=\frac{\bar{A}-A_0}{\bar{A}}=1-\frac{A_0}{\bar{A}} \ .
\label{eq:Appendix_1}
\end{eqnarray}
In case of $\delta t=4$ and 8\,ns, peaks A$_1$ and A$_1'$ are overlapping
with the adjacent cluster. Hence, the visibility is expressed by
\begin{eqnarray}
V=\frac{2\tilde{A}/3-A_0}{2\tilde{A}/3}=1-\frac{A_0}{2\tilde{A}/3} \ ,
\label{eq:Appendix_2}
\end{eqnarray}
with $\tilde{A}$ being the mean value of A$_1$ and A$_1'$ and their related
overlapping peaks. In case of $\delta t=4\,$ns, A$_1$ and A$_1'$ overlap with the
nearest neighbor cluster B$_2$ and B$_2'$. For $\delta t=8\,$ns, the
overlapping peaks stem from C$_2$ and C$_2'$ as seen in
Fig.~\ref{fig:fig_Appendix_1}.
To reduce the statistical error of $\bar{A}$ and $\tilde{A}$, instead of taking only A$_1$, A$_1'$ and their overlapping peak areas into account, we finally averaged over the peak areas for all clusters at $\tau\neq0$, to infer a more precise normalization of the data.
For the pulse separation $\delta t=T=12.5\,$ns, the visibility is determined by
\begin{eqnarray}
V=\frac{\bar{A}_S/2-A_0}{\bar{A}_S/2}=1-\frac{A_0}{\bar{A}_S/2} \ ,
\label{eq:Appendix_3}
\end{eqnarray}
where A$_0$ is the area of the peak at $\tau=0$ and $\bar{A}_S$ corresponds to
the mean value of the side peaks with $\left|\tau\right|>12.5\,$ns.\\
\\
\textbf{Michelson-interferometer measurements:}
\begin{figure}[b]
\includegraphics[width=1\linewidth]{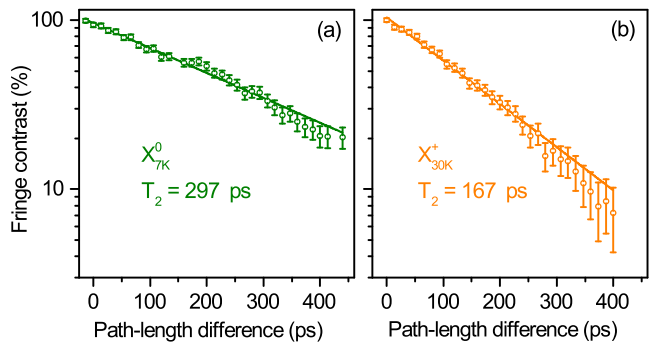}
\caption{\label{fig:fig_Appendix_2} First-order correlation measurements obtained with a fiber-based Michelson-interferometer. Solid lines represent exponential fits to the data for (a) X$^0$ at 7\,K and (b) X$^+$ at 30\,K under p-shell excitation at $\lambda=909\,$nm.}
\end{figure}
We determine the coherence time $T_{2}$ of the charged X$^+$ and neutral X$^0$ exciton state under p-shell excitation at $\lambda=910\,$nm from first-order autocorrelation-measurements $g^{(1)}(\tau)$. Spectrally filtered photons were coupled into a fiber-based Michelson-interferometer to obtain the fringe contrast as a function of the path-length difference. Fitting the data with an exponential decay yields a coherence time of $T_{2}=(291\pm6)\,$ps for the X$^0$ at 7\,K and $T_{2}=(167\pm3)\,$ps for X$^+$ at 30\,K as displayed in Fig.~\ref{fig:fig_Appendix_2}. In both cases the coherence time is significantly below the values of 692\,ps and 431\,ps for X$^0_{7K}$ and X$^+_{30K}$, respectively, obtained by fitting Eq.~3 of the maintext to the experimental data in Fig.~4 of the main text (cf. Table~I in main text) and considering the limit $\delta t\rightarrow\infty$ ($\Gamma^\prime=\Gamma_0^\prime$).
\\
\\
\\
\\
\\
\\
\\
\begin{widetext}
\textbf{Theory:}
In this section, we provide a more detailed derivation of the formula for the 
HOM visibility in Eq.~3 of the maintext. The derivation follows the method
presented by Bylander et al.~\cite{Bylander2003}:
\begin{align}
V(\delta t,\tau_c,T) =& 
1
-
\frac{\Gamma^\prime_0 (1-e^{-\left(\frac{\delta t }{\tau_c}\right)^2}) +
\gamma(T)}{\Gamma^\prime_0
(1-e^{-\left(\frac{\delta t }{\tau_c}\right)^2}) + \gamma(T) + \Gamma}
=
\frac{\Gamma}{\Gamma^\prime_0
(1-e^{-\left(\frac{\delta t }{\tau_c}\right)^2}) + \gamma(T) + \Gamma} \ .
\label{eq:visibility}
\end{align}
Mainly three dephasing/relaxation processes are present in the experiment:
the radiative dephasing $ \Gamma $, the phonon-induced pure dephasing $ \gamma 
$, and the
dephasing due to spectral diffusion $\Gamma^\prime:=\Gamma_0^\prime(1-e^{-(\delta t/\tau_c)^2})$.
\\ \ \\
The total Hamiltonian includes the classical excitation field, the quantized
light field, and the QD. It reads:
\begin{align}
H/\hbar =& 
\left( \omega_e + F(t) \right) \sigma_{ee} 
+ \Omega(t) \left( e^{-i\omega_pt} \sigma_{eg} + e^{+i\omega_pt} \sigma_{ge}
\right)
+ \int_0^\infty d\omega \left( \omega \ c^\dg_\omega c^\ndg_\omega + g_\omega \
c^\dg_\omega \sigma_{ge} + g_\omega^* \sigma_{eg} c^\ndg_\omega \right) \ ,
\end{align}
where we approximated the QD as a two-level system with the
ground $ \ket{g} $ and excited state $ \ket{e} $ and respective lowering and
raising operators defined by $ \sigma_{ij} := \ket{i}\bra{j}$. 
The transition energy between the excited and ground state is denoted by $
\omega_e$, where we set the ground state energy to zero.
To adress dephasing, we apply the commonly used phenomenological 
dephasing description by including a general stochastic force $F(t)$ (here not
specified, see discussion in the paper).
Further, we assume a classical light field with amplitude $\Omega(t)$ in
resonance with the transition energy $ \omega_p=\omega_e $ and a quantized
light
field with annihilation and creation operator $c^\ndg_\omega,c^\dg_\omega$ as well as a
light-matter coupling strength $ g_\omega $, which is assumed to depend only
weakly on frequency $g_\omega \approx g$. 
In the following, we need to distinguish between the photons, which take the
longer route to the final beam splitter to compensate for the earlier emission
process, and those, which reach the beam splitter via the short route.
We distinguish the photons of both channels with the labels: $\omega_L$ for long and $\omega_S$ for
short, i.e. the photons are distinguishable via their spatial travelling
direction until they superpose at the final beam splitter. 
The Hamiltonian is applied to the total wave function, being restricted by the
experiment to the two-photon subspace:
\begin{align}
\notag
\ket{\Psi(t)} =& c_g(t) \ket{g} 
+ c_e(t) \ket{e} 
+ \int d\omega_L \ c_{\omega_L}(t) \ket{g,1_{\omega_L}} 
+ \int d\omega_L \ c^e_{\omega_L}(t) \ket{e,1_{\omega_L}} \\
&\ + \iint d\omega_L d\omega_S \ c_{\omega_S\omega_L}(t)
\ket{g,1_{\omega_L},1_{\omega_S}} \ ,
\end{align}
Here, we assume that the first emitted photon cannot interact with the QD 
a second time, i.e. after the first excitation and subsequent first photon
emission.
Therefore, a state, such as $ \ket{g,2_{\omega_L},0_{\omega_S}} $ is not 
taken into account, provided that the excitation process is faster than the 
emission time
scale. 
Switching into the interaction picture and assuming that the p-shell excitation
is fast compared to any quantum optical emission dynamics, we start with an initial
condition $ c_e(0)=1 $ and solve the Wigner-Weisskopf problem:
\begin{align}
\partial_t c_e 
= & -i g \int d\omega \ e^{-i(\omega-\omega_e)t +i \phi(t)} c_\omega \\
\partial_t c_\omega 
= & -i \ g \ e^{i(\omega-\omega_e)t -i \phi(t)} c_e\ .
\end{align}
Integrating the latter equation formally and plug them into the first
equation, one yields a simple relaxation dynamics for the excited state, and the
corresponding photon wave package form:
\begin{align}
\partial_t c_e = & - g^2 \pi c_e(t) \rightarrow c_e(t) =  c_e(0)
e^{-\Gamma t} \\
c_\omega(t)  
= & -i \ g \ \int_0^t dt^\prime e^{i(\omega-\omega_e)t^\prime -i \phi(t^\prime)}
c_e(0) e^{-\Gamma t^\prime} \ ,
\end{align}
using the abbreviation $ \Gamma = g^2 \pi$ and including the frequency shift
into the resonance condition. Note, we restricted our analysis to a
one-dimensional problem.
This is in accordance to the experiment, where the fiber and the optical setup
allow for such a treatment.
\\ \ \\
For the subsequent excitation pulse, we assume $ |c_e(t)|^2 \approx 0$,
with $ t>\delta t$ and $ \delta t $ the temporal distance between the two 
excitation pulses.
Since the first photon wave package cannot interact with the QD 
anymore, the wave function between the photonic and electronic part factorizes 
and the same calculation as before can be applied. 
The two-photon wave function reads:
\begin{align} \notag
\ket{\Psi(t)} = & 
\left[
-i g \
\int_0^\infty d\omega_S 
\int_{\delta t}^t dt_S e^{i(\omega_S-\omega_e)t_S-i\phi_{\delta t}(t_S)-\Gamma t_S} 
\ c^\dg_{\omega_S} \ket{0_{\omega_S}} \right] \\
\ &
\otimes
\left[
-i g \
\int_0^\infty d\omega_L 
\int_0^t dt_L e^{i(\omega_L-\omega_e)t_L-i\phi_0(t_L)-\Gamma t_L}
c^\dg_{\omega_L} \ket{0_{\omega_L}}  \right]\ .
\end{align}
Note, the difference in the lower limit of the integrals ($0,\delta t$) 
and in the integrated noise signals
\begin{align}
\phi_{t_1}(t) =& \int_{t_1}^t dt^\prime F(t^\prime) \ .
\end{align}
Given this wave function, we can calculate the observables of the experiment, 
as discussed in the following.

The Hong-Ou-Mandel effect leads to a vanishing two-photon-correlation for a 
pair of indistinguishable photons as both photons travelling either via the
transmission or reflection path through the beam splitter.
The quantity of interest is this two-photon correlation 
g$^{(2)}(t_D,t_D+\tau)$ between photons measured during the time $t_D$ on detector $A$ and $B$ with 
a delay of $\tau$: 
\begin{align}
g^{(2)}(t_D,t_D+\tau) =& \frac{\bra{\Psi(t)} E^{(-)}_A(t_D) E^{(-)}_B(t_D+\tau)
E^{(+)}_B(t_D+\tau) E^{(+)}_A(t_D)\ket{\Psi(t_D)} }{\bra{\Psi(t)}
E^{(-)}_B(t_D+\tau)
E^{(+)}_B(t_D+\tau)\ket{\Psi(t)} \bra{\Psi(t)} E^{(-)}_A(t_D)
E^{(+)}_A(t_D)\ket{\Psi(t)} } \ .
\end{align}
The electric fields $ E_A $ and $ E_B $ are calculated via the incoming fields 
of the long and short fiber and the transmission and reflection coefficients:
\begin{align}
E^{(\pm)}_A =& \sqrt{T} E^{(\pm)}_S + \sqrt{R} E^{(\pm)}_L  \\
E^{(\pm)}_B =& \sqrt{T} E^{(\pm)}_L - \sqrt{R} E^{(\pm)}_S \ . 
\end{align}
For the two-photon correlation, we need to calculate the ket:
\begin{align} \notag
E^{(+)}(t_D+\tau)_B E^{(+)}_A(t_D) \ket{\psi(t)} =&
\left[
T 
E^{(+)}_S(t_D+\tau) E^{(+)}_L(t_D) 
-
R
E^{(+)}_L(t_D+\tau) E^{(+)}_S(t_D) \right] \ket{\psi(t)} \ .
\end{align}
Here, we omitted all contributions, where two photons in the long or short
channel are needed.
Using the commutation relations $ 
[c^\ndg_{\omega_i},c^\dg_{\omega^\prime_j}]=\delta_{ij}
\delta(\omega-\omega^\prime) $, we can evaluate
the ket further and yield:
\begin{align}
\notag
& E^{(+)}_B(t_D+\tau) E^{(+)}_A(t_D)\ket{\Psi(t)} 
 \ \\
&= 
[-ig]^2
\iint d\omega_2 d\omega_1 
\left(
T
e^{-i\omega_2(t_D+\tau)} 
e^{-i\omega_1t_D}
-
R
e^{-i\omega_1(t_D+\tau)} 
e^{-i\omega_2t_D} 
\right) \\ \notag 
&
\qquad \qquad \int_0^t dt^{\prime\prime}
e^{i(\omega_1-\omega_e)t^{\prime\prime} -i
\phi(t^{\prime\prime})} e^{-\Gamma t^{\prime\prime}}
\int_T^t dt^\prime e^{i(\omega_2-\omega_e)t^\prime -i
\phi_T(t^\prime)} e^{-\Gamma t^\prime} 
\ket{g,0,0} \ .
\end{align}
Integrating over the frequencies and considering the long-time limit $ 
t\rightarrow \infty$, we yield for the unnormalized two-photon correlation:
\begin{align}
\notag
& E^{(+)}_B(t_D+\tau) E^{(+)}_A(t_D)\ket{\Psi(t)} 
 \ \\
&= 
-g^2\pi^2
\left[
T
e^{-i\phi(t_D+\tau)-i\phi_{\delta t}(t_D)}
-
R
e^{-i\phi_{\delta t}(t_D+\tau)-i\phi_{\delta t}(t_D)}
\right] 
e^{-i\omega_e(2t_D+\tau)-\Gamma(2t_D+\tau)}
\ket{g,0,0} \ . \label{eq:ket_2photon}
\end{align}
The unnormalized two-photon correlation $G^{(2)}(t,\tau)$ is the product of 
Eq.~\eqref{eq:ket_2photon} and its conjugate:
\begin{align}
& G^{(2)}(t_D,\tau) = 
g^4\pi^4
e^{-\Gamma(2t_D+\tau)} 
\left[
T^2 + R^2
-2RT \ \text{Re}\left[\left\langle
e^{-i\phi(t_D+\tau)-i\phi_{\delta t}(t_D)+i\phi_{\delta t}(t_D+\tau)+i\phi(t_D)} 
\right\rangle\right]
\right] \ , \label{eq:corr_two_photon} 
\end{align}
where we already wrote the statistical averaging into the formula by $ \langle 
\cdot \rangle $.
At this point, we need to specify the noise correlations to evaluate this
expression further.
To evaluate the stochastic forces, we need to average via a Gaussian 
random number distribution $ \left\langle\left\langle \cdot
\right\rangle\right\rangle $, where all higher moments can be expressed by the second-order correlation \cite{Gardiner2004}.
Eq.~\eqref{eq:corr_two_photon} is still very general in terms of dephasing 
processes and can be evaluated for Markovian- and non-Markovian noise 
correlations.
To include dephasing, we employ the working horse of the phenomenological 
dephasing description by including a general stochastic force
$F(t) = P(t) + D(t)$ with a phonon-induced dephasing
($\delta$-correlated white noise) $P(t)$ and a spectral diffusion $D(t)$
component (colored noise), both shifting the transition energy of the QD.
First, we assume that the phonon-induced dephasing and the dephasing stemming
from the spectral diffusion in the material are independent of each other.
Therefore, we can neglect correlations between $ D(t) $ and $ P(t) $ in the
cumulant expansion.
Here, we restrict our investigation to the zero-phonon line broadening 
mechanism \cite{Borri2001}.
A possible source for such a dephasing mechanism is the quadratic 
interaction with longitudinal acoustical phonons, which gives rise to a 
temperature-dependent broadening \cite{Foerstner2003,Muljarov2004}.
We also neglect contributions from highly non-Markovian phonon-sidebands, which
also effect the indistiguishability, e.g. in cQED setups \cite{Kaer2013,Kaer2013a}, and described often with the 
independent Boson model \cite{Stock2011b} or Feynman path 
integrals \cite{Vagov2011}.
Here, we employ the simplest possible model for such a dephasing by assuming a 
Markovian process $\delta-$correlated in time, i.e. as white noise 
\cite{Gardiner2004,Santori2009}.

\begin{align}
\left\langle\left\langle  
\phi^P_{t_1}(t_2)\phi^P_{t_1}(t_4)
\right\rangle\right\rangle  
=
\int_{t_1}^{t_2}
dt
\int_{t_3}^{t_4}
dt^\prime 
\left\langle P(t) P(t^\prime) \right\rangle
=& \gamma \left( \text{min}[t_2,t_4] - \text{max}[t_1,t_3] \right) \ .
\end{align}
The phonon-induced dephasing is highly temperature-dependent and limits the 
absolute value of the indistinguishability, independent of the temporal
distance of the excitation pulses $ \delta t$.
In contrast to the phonon-induced dephasing, the spectral diffusion includes
a strong dependence on the pulse distance.
We include this dependence as a finite memory-effect with specific 
correlation time $ \tau_c$: 
\begin{align}
\left\langle\left\langle  
\phi^D_{t_1}(t_2) \phi^D_{t_3}(t_4)
\right\rangle\right\rangle  
=& 
\int_{t_1}^{t_2}
dt
\int_{t_3}^{t_4}
dt^\prime 
\left\langle D(t) D(t^\prime) \right\rangle
= \Gamma^\prime_0 \
e^{-\frac{(t_1-t_3)^2}{\tau_c^2}}
\left( \text{min}[t_2,t_4] - \text{max}[t_1,t_3] \right) \ .
\end{align}
These kinds of noise correlations stem from a non-Markovian low-frequency 
noise \cite{Galperin2006,Laikhtman1985,Eberly1984} and 
show plateau-like behavior for temporal pulse distances sufficiently
short in comparison to the memory depth, i.e. for $ \delta t \ll \tau_c $ the
effect of spectral diffusion becomes negligible and only the phonon-induced
dephasing limits the absolute value of the visibility.
\\ \ \\ 
The unnormalized two-photon correlation then reads:
\begin{align}
 G^{(2)}(t_D,\tau) =& g^4\pi^4e^{-2\Gamma t_D} \left[ T^2 e^{-\Gamma \tau}  + R^2 e^{-\Gamma \tau} -2RT \ e^{-(\gamma^\prime+\Gamma) \tau} \right], \label{eq:corr_two_photon_after_averaging} \\
 \gamma^\prime =& \Gamma^\prime_0 \left( 1-\exp[-(\delta t/\tau_c)^2] \right) + 
\gamma \ ,
\end{align}
with $\delta t$ the temporal pulse distance, $ \Gamma^\prime_0 $ the spectral 
diffusion constant, and $ \gamma $ the phonon-induced dephasing.
As the measured quantity is the time-integrated photon correlation, integrating 
with respect to $ t_D $ and $ \tau $ yields, using $ R+T=1$:
\begin{align}
& \bar g^{(2)} = \frac{2}{\pi^2(T^2+R^2)} \int_0^\infty\int_0^\infty 
\text{d}\tau \text{d}t \ G^{(2)}(t,\tau)
=
1 - \frac{2RT}{1-2RT} \frac{\Gamma}{\gamma^\prime + \Gamma} \ .
\end{align}
The visibility can be expressed via the normalized two-photon-correlation
\begin{align}
V=& 1 - \bar g^{(2)} 
= 
\frac{2RT}{1-2RT} \left[ 1 - \frac{\gamma^\prime}{\gamma^\prime + \Gamma}  
\right] \ .
\end{align}
With these equations at hand, we are now able explicitly formulate
the dependence of the visibility on the pulse separation $ \delta t $, the 
pure dephasing $ \gamma $, and the diffusion constant $ \Gamma^\prime $:
\begin{align}
V=& 
\frac{\Gamma}{\Gamma^\prime_0
(1-\exp[-(\delta t/\tau_c)^2])+\gamma(T) + \Gamma} \ , \label{eq:visib_final}
\end{align}
where a balanced beamsplitter ($ T = R = 1/2 $) was assumed.
Thus, for vanishing phonon-induced dephasing and spectral diffusion, the
visibility is $1$, i.e. the photons are only Fourier-transform-limited and
coalesce at the beamsplitter into a perfect coherent two-photon state.
If the phonon-induced dephasing is stronger than other dephasing and relaxation
processes $ \gamma \gg \Gamma,\Gamma^\prime$, the visibility becomes small,
which is typically seen in the high temperature limit.
At low temperatures, the phonon-induced dephasing is small and the spectral
diffusion with a finite-memory depth dictates the functional form of the
visibility for different pulse distances.
\\ \ \\
To approximate the temperature dependence of the visibility, we employ the
Markovian approximation for phonon-induced pure dephasing processes, where 
the dephasing is proportional to the square of the phonon number \cite{Carmichael1999}:
\begin{align}
\gamma(T) &= \gamma_0 \ \bar n(T) 
\left[ \bar n(T) + 1 \right] \ ,
\end{align}
where we have averaged over the frequency and approximated the expression via
an effective phonon number depending on the temperature via the Bose-Einstein 
distribution for the effective phonon mode.
The following formula is employed to underline the experimentally observed behavior qualitatively:
\begin{align}
\bar n(T) =&
\left[ \exp\left[\frac{\alpha}{T} \right] - 1 \right]^{-1} \ .
\end{align}
To fit the curve in Fig.~5\,(d) in the maintext, we adjust the parameters $ \gamma_0 $ and $ \alpha $.
For illustrating purposes, we normalized the other dephasing contributions to one:
\begin{align}
V(T)=& 
\frac{\Gamma}{\Gamma^\prime_0
(1-\exp[-(\delta t/\tau_c)^2])+\gamma(T) + \Gamma}
\approx \frac{1}{1+\gamma_0 \ \bar n(\alpha,T) 
\left[ \bar n(\alpha,T) + 1 \right]} \ . \label{eq:visib_T}
\end{align}
The fit presented in Fig.~5 of the maintext, according to this formula, was performed with $ \alpha=\hbar\bar\omega/k_B=44\,$K and $ \gamma_0 = 3.75 $.
\end{widetext}

\bibliography{Thoma_et_al}

\begin{thebibliography}{46}%
\makeatletter
\providecommand \@ifxundefined [1]{%
 \@ifx{#1\undefined}
}%
\providecommand \@ifnum [1]{%
 \ifnum #1\expandafter \@firstoftwo
 \else \expandafter \@secondoftwo
 \fi
}%
\providecommand \@ifx [1]{%
 \ifx #1\expandafter \@firstoftwo
 \else \expandafter \@secondoftwo
 \fi
}%
\providecommand \natexlab [1]{#1}%
\providecommand \enquote  [1]{``#1''}%
\providecommand \bibnamefont  [1]{#1}%
\providecommand \bibfnamefont [1]{#1}%
\providecommand \citenamefont [1]{#1}%
\providecommand \href@noop [0]{\@secondoftwo}%
\providecommand \href [0]{\begingroup \@sanitize@url \@href}%
\providecommand \@href[1]{\@@startlink{#1}\@@href}%
\providecommand \@@href[1]{\endgroup#1\@@endlink}%
\providecommand \@sanitize@url [0]{\catcode `\\12\catcode `\$12\catcode
  `\&12\catcode `\#12\catcode `\^12\catcode `\_12\catcode `\%12\relax}%
\providecommand \@@startlink[1]{}%
\providecommand \@@endlink[0]{}%
\providecommand \url  [0]{\begingroup\@sanitize@url \@url }%
\providecommand \@url [1]{\endgroup\@href {#1}{\urlprefix }}%
\providecommand \urlprefix  [0]{URL }%
\providecommand \Eprint [0]{\href }%
\providecommand \doibase [0]{http://dx.doi.org/}%
\providecommand \selectlanguage [0]{\@gobble}%
\providecommand \bibinfo  [0]{\@secondoftwo}%
\providecommand \bibfield  [0]{\@secondoftwo}%
\providecommand \translation [1]{[#1]}%
\providecommand \BibitemOpen [0]{}%
\providecommand \bibitemStop [0]{}%
\providecommand \bibitemNoStop [0]{.\EOS\space}%
\providecommand \EOS [0]{\spacefactor3000\relax}%
\providecommand \BibitemShut  [1]{\csname bibitem#1\endcsname}%
\let\auto@bib@innerbib\@empty
\bibitem [{\citenamefont {Knill}\ \emph {et~al.}(2001)\citenamefont {Knill},
  \citenamefont {Laflamme},\ and\ \citenamefont {Milburn}}]{Knill2001}%
  \BibitemOpen
  \bibfield  {author} {\bibinfo {author} {\bibfnamefont {E.}~\bibnamefont
  {Knill}}, \bibinfo {author} {\bibfnamefont {R.}~\bibnamefont {Laflamme}}, \
  and\ \bibinfo {author} {\bibfnamefont {G.~J.}\ \bibnamefont {Milburn}},\
  }\href {http://dx.doi.org/10.1038/35051009} {\bibfield  {journal} {\bibinfo
  {journal} {Nature}\ }\textbf {\bibinfo {volume} {409}},\ \bibinfo {pages}
  {46} (\bibinfo {year} {2001})}\BibitemShut {NoStop}%
\bibitem [{\citenamefont {Kiraz}\ \emph {et~al.}(2004)\citenamefont {Kiraz},
  \citenamefont {Atat\"ure},\ and\ \citenamefont {Imamo\ifmmode~\breve{g}\else
  \u{g}\fi{}lu}}]{Kiraz2004}%
  \BibitemOpen
  \bibfield  {author} {\bibinfo {author} {\bibfnamefont {A.}~\bibnamefont
  {Kiraz}}, \bibinfo {author} {\bibfnamefont {M.}~\bibnamefont {Atat\"ure}}, \
  and\ \bibinfo {author} {\bibfnamefont {A.}~\bibnamefont
  {Imamo\ifmmode~\breve{g}\else \u{g}\fi{}lu}},\ }\href {\doibase
  10.1103/PhysRevA.69.032305} {\bibfield  {journal} {\bibinfo  {journal} {Phys.
  Rev. A}\ }\textbf {\bibinfo {volume} {69}},\ \bibinfo {pages} {032305}
  (\bibinfo {year} {2004})}\BibitemShut {NoStop}%
\bibitem [{\citenamefont {Kok}\ \emph {et~al.}(2007)\citenamefont {Kok},
  \citenamefont {Munro}, \citenamefont {Nemoto}, \citenamefont {Ralph},
  \citenamefont {Dowling},\ and\ \citenamefont {Milburn}}]{Kok2007}%
  \BibitemOpen
  \bibfield  {author} {\bibinfo {author} {\bibfnamefont {P.}~\bibnamefont
  {Kok}}, \bibinfo {author} {\bibfnamefont {W.~J.}\ \bibnamefont {Munro}},
  \bibinfo {author} {\bibfnamefont {K.}~\bibnamefont {Nemoto}}, \bibinfo
  {author} {\bibfnamefont {T.~C.}\ \bibnamefont {Ralph}}, \bibinfo {author}
  {\bibfnamefont {J.~P.}\ \bibnamefont {Dowling}}, \ and\ \bibinfo {author}
  {\bibfnamefont {G.~J.}\ \bibnamefont {Milburn}},\ }\href {\doibase
  10.1103/RevModPhys.79.135} {\bibfield  {journal} {\bibinfo  {journal} {Rev.
  Mod. Phys.}\ }\textbf {\bibinfo {volume} {79}},\ \bibinfo {pages} {135}
  (\bibinfo {year} {2007})}\BibitemShut {NoStop}%
\bibitem [{\citenamefont {Gisin}\ and\ \citenamefont {Thew}(2007)}]{Gisin2007}%
  \BibitemOpen
  \bibfield  {author} {\bibinfo {author} {\bibfnamefont {N.}~\bibnamefont
  {Gisin}}\ and\ \bibinfo {author} {\bibfnamefont {R.}~\bibnamefont {Thew}},\
  }\href {http://dx.doi.org/10.1038/nphoton.2007.22} {\bibfield  {journal}
  {\bibinfo  {journal} {Nature Photon.}\ }\textbf {\bibinfo {volume} {1}},\
  \bibinfo {pages} {165} (\bibinfo {year} {2007})}\BibitemShut {NoStop}%
\bibitem [{\citenamefont {Kimble}(2008)}]{Kimble2008}%
  \BibitemOpen
  \bibfield  {author} {\bibinfo {author} {\bibfnamefont {H.~J.}\ \bibnamefont
  {Kimble}},\ }\href {\doibase 10.1038/nature07127} {\bibfield  {journal}
  {\bibinfo  {journal} {Nature}\ }\textbf {\bibinfo {volume} {453}},\ \bibinfo
  {pages} {1023} (\bibinfo {year} {2008})}\BibitemShut {NoStop}%
\bibitem [{\citenamefont {Michler}\ \emph {et~al.}(2000)\citenamefont
  {Michler}, \citenamefont {Kiraz}, \citenamefont {Becher}, \citenamefont
  {Schoenfeld}, \citenamefont {Petroff}, \citenamefont {Zhang}, \citenamefont
  {Hu},\ and\ \citenamefont {Imamo\u{g}lu}}]{Michler2000}%
  \BibitemOpen
  \bibfield  {author} {\bibinfo {author} {\bibfnamefont {P.}~\bibnamefont
  {Michler}}, \bibinfo {author} {\bibfnamefont {A.}~\bibnamefont {Kiraz}},
  \bibinfo {author} {\bibfnamefont {C.}~\bibnamefont {Becher}}, \bibinfo
  {author} {\bibfnamefont {W.~V.}\ \bibnamefont {Schoenfeld}}, \bibinfo
  {author} {\bibfnamefont {P.~M.}\ \bibnamefont {Petroff}}, \bibinfo {author}
  {\bibfnamefont {L.}~\bibnamefont {Zhang}}, \bibinfo {author} {\bibfnamefont
  {E.}~\bibnamefont {Hu}}, \ and\ \bibinfo {author} {\bibfnamefont
  {A.}~\bibnamefont {Imamo\u{g}lu}},\ }\href@noop {} {\bibfield  {journal}
  {\bibinfo  {journal} {Science}\ }\textbf {\bibinfo {volume} {290}},\ \bibinfo
  {pages} {2282} (\bibinfo {year} {2000})}\BibitemShut {NoStop}%
\bibitem [{\citenamefont {Santori}\ \emph {et~al.}(2002)\citenamefont
  {Santori}, \citenamefont {Fattal}, \citenamefont {Vu\v{c}kovi\'{c}},
  \citenamefont {Solomon},\ and\ \citenamefont {Yamamoto}}]{Santori2002b}%
  \BibitemOpen
  \bibfield  {author} {\bibinfo {author} {\bibfnamefont {C.}~\bibnamefont
  {Santori}}, \bibinfo {author} {\bibfnamefont {D.}~\bibnamefont {Fattal}},
  \bibinfo {author} {\bibfnamefont {J.}~\bibnamefont {Vu\v{c}kovi\'{c}}},
  \bibinfo {author} {\bibfnamefont {G.~S.}\ \bibnamefont {Solomon}}, \ and\
  \bibinfo {author} {\bibfnamefont {Y.}~\bibnamefont {Yamamoto}},\ }\href
  {\doibase 10.1038/nature01086} {\bibfield  {journal} {\bibinfo  {journal}
  {Nature}\ }\textbf {\bibinfo {volume} {419}},\ \bibinfo {pages} {594}
  (\bibinfo {year} {2002})}\BibitemShut {NoStop}%
\bibitem [{\citenamefont {Patel}\ \emph {et~al.}(2008)\citenamefont {Patel},
  \citenamefont {Bennett}, \citenamefont {Cooper}, \citenamefont {Atkinson},
  \citenamefont {Nicoll}, \citenamefont {Ritchie},\ and\ \citenamefont
  {Shields}}]{Patel2008}%
  \BibitemOpen
  \bibfield  {author} {\bibinfo {author} {\bibfnamefont {R.~B.}\ \bibnamefont
  {Patel}}, \bibinfo {author} {\bibfnamefont {A.~J.}\ \bibnamefont {Bennett}},
  \bibinfo {author} {\bibfnamefont {K.}~\bibnamefont {Cooper}}, \bibinfo
  {author} {\bibfnamefont {P.}~\bibnamefont {Atkinson}}, \bibinfo {author}
  {\bibfnamefont {C.~A.}\ \bibnamefont {Nicoll}}, \bibinfo {author}
  {\bibfnamefont {D.~A.}\ \bibnamefont {Ritchie}}, \ and\ \bibinfo {author}
  {\bibfnamefont {A.~J.}\ \bibnamefont {Shields}},\ }\href {\doibase
  10.1103/PhysRevLett.100.207405} {\bibfield  {journal} {\bibinfo  {journal}
  {Phys. Rev. Lett.}\ }\textbf {\bibinfo {volume} {100}},\ \bibinfo {pages}
  {207405} (\bibinfo {year} {2008})}\BibitemShut {NoStop}%
\bibitem [{\citenamefont {Ates}\ \emph {et~al.}(2009)\citenamefont {Ates},
  \citenamefont {Ulrich}, \citenamefont {Ulhaq}, \citenamefont {Reitzenstein},
  \citenamefont {Loffler}, \citenamefont {Hofling}, \citenamefont {Forchel},\
  and\ \citenamefont {Michler}}]{Ates2009}%
  \BibitemOpen
  \bibfield  {author} {\bibinfo {author} {\bibfnamefont {S.}~\bibnamefont
  {Ates}}, \bibinfo {author} {\bibfnamefont {S.~M.}\ \bibnamefont {Ulrich}},
  \bibinfo {author} {\bibfnamefont {A.}~\bibnamefont {Ulhaq}}, \bibinfo
  {author} {\bibfnamefont {S.}~\bibnamefont {Reitzenstein}}, \bibinfo {author}
  {\bibfnamefont {A.}~\bibnamefont {Loffler}}, \bibinfo {author} {\bibfnamefont
  {S.}~\bibnamefont {Hofling}}, \bibinfo {author} {\bibfnamefont
  {A.}~\bibnamefont {Forchel}}, \ and\ \bibinfo {author} {\bibfnamefont
  {P.}~\bibnamefont {Michler}},\ }\href {\doibase 10.1038/nphoton.2009.215}
  {\bibfield  {journal} {\bibinfo  {journal} {Nature Photon.}\ }\textbf
  {\bibinfo {volume} {3}},\ \bibinfo {pages} {724} (\bibinfo {year}
  {2009})}\BibitemShut {NoStop}%
\bibitem [{\citenamefont {Wei}\ \emph {et~al.}(2014)\citenamefont {Wei},
  \citenamefont {He}, \citenamefont {Chen}, \citenamefont {Hu}, \citenamefont
  {He}, \citenamefont {Wu}, \citenamefont {Schneider}, \citenamefont {Kamp},
  \citenamefont {Höfling}, \citenamefont {Lu},\ and\ \citenamefont
  {Pan}}]{Wei2014}%
  \BibitemOpen
  \bibfield  {author} {\bibinfo {author} {\bibfnamefont {Y.-J.}\ \bibnamefont
  {Wei}}, \bibinfo {author} {\bibfnamefont {Y.-M.}\ \bibnamefont {He}},
  \bibinfo {author} {\bibfnamefont {M.-C.}\ \bibnamefont {Chen}}, \bibinfo
  {author} {\bibfnamefont {Y.-N.}\ \bibnamefont {Hu}}, \bibinfo {author}
  {\bibfnamefont {Y.}~\bibnamefont {He}}, \bibinfo {author} {\bibfnamefont
  {D.}~\bibnamefont {Wu}}, \bibinfo {author} {\bibfnamefont {C.}~\bibnamefont
  {Schneider}}, \bibinfo {author} {\bibfnamefont {M.}~\bibnamefont {Kamp}},
  \bibinfo {author} {\bibfnamefont {S.}~\bibnamefont {Höfling}}, \bibinfo
  {author} {\bibfnamefont {C.-Y.}\ \bibnamefont {Lu}}, \ and\ \bibinfo {author}
  {\bibfnamefont {J.-W.}\ \bibnamefont {Pan}},\ }\href {\doibase
  10.1021/nl503081n} {\bibfield  {journal} {\bibinfo  {journal} {Nano Lett.}\
  }\textbf {\bibinfo {volume} {14}},\ \bibinfo {pages} {6515} (\bibinfo {year}
  {2014})}\BibitemShut {NoStop}%
\bibitem [{\citenamefont {Santori}\ \emph {et~al.}(2004)\citenamefont
  {Santori}, \citenamefont {Fattal}, \citenamefont {Vuckovic}, \citenamefont
  {Solomon},\ and\ \citenamefont {Yamamoto}}]{Santori2004}%
  \BibitemOpen
  \bibfield  {author} {\bibinfo {author} {\bibfnamefont {C.}~\bibnamefont
  {Santori}}, \bibinfo {author} {\bibfnamefont {D.}~\bibnamefont {Fattal}},
  \bibinfo {author} {\bibfnamefont {J.}~\bibnamefont {Vuckovic}}, \bibinfo
  {author} {\bibfnamefont {G.~S.}\ \bibnamefont {Solomon}}, \ and\ \bibinfo
  {author} {\bibfnamefont {Y.}~\bibnamefont {Yamamoto}},\ }\href
  {http://stacks.iop.org/1367-2630/6/i=1/a=089} {\bibfield  {journal} {\bibinfo
   {journal} {New J. Phys.}\ }\textbf {\bibinfo {volume} {6}},\ \bibinfo
  {pages} {89} (\bibinfo {year} {2004})}\BibitemShut {NoStop}%
\bibitem [{\citenamefont {Varoutsis}\ \emph {et~al.}(2005)\citenamefont
  {Varoutsis}, \citenamefont {Laurent}, \citenamefont {Kramper}, \citenamefont
  {Lema\^{i}tre}, \citenamefont {Sagnes}, \citenamefont {Robert-Philip},\ and\
  \citenamefont {Abram}}]{Varoutsis2005}%
  \BibitemOpen
  \bibfield  {author} {\bibinfo {author} {\bibfnamefont {S.}~\bibnamefont
  {Varoutsis}}, \bibinfo {author} {\bibfnamefont {S.}~\bibnamefont {Laurent}},
  \bibinfo {author} {\bibfnamefont {P.}~\bibnamefont {Kramper}}, \bibinfo
  {author} {\bibfnamefont {A.}~\bibnamefont {Lema\^{i}tre}}, \bibinfo {author}
  {\bibfnamefont {I.}~\bibnamefont {Sagnes}}, \bibinfo {author} {\bibfnamefont
  {I.}~\bibnamefont {Robert-Philip}}, \ and\ \bibinfo {author} {\bibfnamefont
  {I.}~\bibnamefont {Abram}},\ }\href {\doibase 10.1103/physrevb.72.041303}
  {\bibfield  {journal} {\bibinfo  {journal} {Physical Review B}\ }\textbf
  {\bibinfo {volume} {72}},\ \bibinfo {pages} {041303} (\bibinfo {year}
  {2005})}\BibitemShut {NoStop}%
\bibitem [{\citenamefont {Gazzano}\ \emph {et~al.}(2013)\citenamefont
  {Gazzano}, \citenamefont {Michaelis~de Vasconcellos}, \citenamefont {Arnold},
  \citenamefont {Nowak}, \citenamefont {Galopin}, \citenamefont {Sagnes},
  \citenamefont {Lanco}, \citenamefont {Lema\^{i}tre},\ and\ \citenamefont
  {Senellart}}]{Gazzano2013}%
  \BibitemOpen
  \bibfield  {author} {\bibinfo {author} {\bibfnamefont {O.}~\bibnamefont
  {Gazzano}}, \bibinfo {author} {\bibfnamefont {S.}~\bibnamefont {Michaelis~de
  Vasconcellos}}, \bibinfo {author} {\bibfnamefont {C.}~\bibnamefont {Arnold}},
  \bibinfo {author} {\bibfnamefont {A.}~\bibnamefont {Nowak}}, \bibinfo
  {author} {\bibfnamefont {E.}~\bibnamefont {Galopin}}, \bibinfo {author}
  {\bibfnamefont {I.}~\bibnamefont {Sagnes}}, \bibinfo {author} {\bibfnamefont
  {L.}~\bibnamefont {Lanco}}, \bibinfo {author} {\bibfnamefont
  {A.}~\bibnamefont {Lema\^{i}tre}}, \ and\ \bibinfo {author} {\bibfnamefont
  {P.}~\bibnamefont {Senellart}},\ }\href
  {http://dx.doi.org/10.1038/ncomms2434} {\bibfield  {journal} {\bibinfo
  {journal} {Nat. Commun.}\ }\textbf {\bibinfo {volume} {4}},\ \bibinfo {pages}
  {1425} (\bibinfo {year} {2013})}\BibitemShut {NoStop}%
\bibitem [{\citenamefont {Gold}\ \emph {et~al.}(2014)\citenamefont {Gold},
  \citenamefont {Thoma}, \citenamefont {Maier}, \citenamefont {Reitzenstein},
  \citenamefont {Schneider}, \citenamefont {H\"ofling},\ and\ \citenamefont
  {Kamp}}]{Gold2014}%
  \BibitemOpen
  \bibfield  {author} {\bibinfo {author} {\bibfnamefont {P.}~\bibnamefont
  {Gold}}, \bibinfo {author} {\bibfnamefont {A.}~\bibnamefont {Thoma}},
  \bibinfo {author} {\bibfnamefont {S.}~\bibnamefont {Maier}}, \bibinfo
  {author} {\bibfnamefont {S.}~\bibnamefont {Reitzenstein}}, \bibinfo {author}
  {\bibfnamefont {C.}~\bibnamefont {Schneider}}, \bibinfo {author}
  {\bibfnamefont {S.}~\bibnamefont {H\"ofling}}, \ and\ \bibinfo {author}
  {\bibfnamefont {M.}~\bibnamefont {Kamp}},\ }\href {\doibase
  10.1103/PhysRevB.89.035313} {\bibfield  {journal} {\bibinfo  {journal} {Phys.
  Rev. B}\ }\textbf {\bibinfo {volume} {89}},\ \bibinfo {pages} {035313}
  (\bibinfo {year} {2014})}\BibitemShut {NoStop}%
\bibitem [{\citenamefont {Hong}\ \emph {et~al.}(1987)\citenamefont {Hong},
  \citenamefont {Ou},\ and\ \citenamefont {Mandel}}]{Hong1987}%
  \BibitemOpen
  \bibfield  {author} {\bibinfo {author} {\bibfnamefont {C.~K.}\ \bibnamefont
  {Hong}}, \bibinfo {author} {\bibfnamefont {Z.~Y.}\ \bibnamefont {Ou}}, \ and\
  \bibinfo {author} {\bibfnamefont {L.}~\bibnamefont {Mandel}},\ }\href
  {\doibase 10.1103/PhysRevLett.59.2044} {\bibfield  {journal} {\bibinfo
  {journal} {Phys. Rev. Lett.}\ }\textbf {\bibinfo {volume} {59}},\ \bibinfo
  {pages} {2044} (\bibinfo {year} {1987})}\BibitemShut {NoStop}%
\bibitem [{\citenamefont {Bennett}\ \emph {et~al.}(2005)\citenamefont
  {Bennett}, \citenamefont {Unitt}, \citenamefont {Shields}, \citenamefont
  {Atkinson},\ and\ \citenamefont {Ritchie}}]{Bennett2005c}%
  \BibitemOpen
  \bibfield  {author} {\bibinfo {author} {\bibfnamefont {A.~J.}\ \bibnamefont
  {Bennett}}, \bibinfo {author} {\bibfnamefont {D.~C.}\ \bibnamefont {Unitt}},
  \bibinfo {author} {\bibfnamefont {A.~J.}\ \bibnamefont {Shields}}, \bibinfo
  {author} {\bibfnamefont {P.}~\bibnamefont {Atkinson}}, \ and\ \bibinfo
  {author} {\bibfnamefont {D.~A.}\ \bibnamefont {Ritchie}},\ }\href {\doibase
  10.1364/opex.13.007772} {\bibfield  {journal} {\bibinfo  {journal} {Opt.
  Express}\ }\textbf {\bibinfo {volume} {13}},\ \bibinfo {pages} {7772}
  (\bibinfo {year} {2005})}\BibitemShut {NoStop}%
\bibitem [{\citenamefont {Wolters}\ \emph {et~al.}(2013)\citenamefont
  {Wolters}, \citenamefont {Sadzak}, \citenamefont {Schell}, \citenamefont
  {Schr\"oder},\ and\ \citenamefont {Benson}}]{Wolters2013}%
  \BibitemOpen
  \bibfield  {author} {\bibinfo {author} {\bibfnamefont {J.}~\bibnamefont
  {Wolters}}, \bibinfo {author} {\bibfnamefont {N.}~\bibnamefont {Sadzak}},
  \bibinfo {author} {\bibfnamefont {A.~W.}\ \bibnamefont {Schell}}, \bibinfo
  {author} {\bibfnamefont {T.}~\bibnamefont {Schr\"oder}}, \ and\ \bibinfo
  {author} {\bibfnamefont {O.}~\bibnamefont {Benson}},\ }\href {\doibase
  10.1103/PhysRevLett.110.027401} {\bibfield  {journal} {\bibinfo  {journal}
  {Phys. Rev. Lett.}\ }\textbf {\bibinfo {volume} {110}},\ \bibinfo {pages}
  {027401} (\bibinfo {year} {2013})}\BibitemShut {NoStop}%
\bibitem [{\citenamefont {Stanley}\ \emph {et~al.}(2014)\citenamefont
  {Stanley}, \citenamefont {Matthiesen}, \citenamefont {Hansom}, \citenamefont
  {Le~Gall}, \citenamefont {Schulte}, \citenamefont {Clarke},\ and\
  \citenamefont {Atatüre}}]{Stanley2014}%
  \BibitemOpen
  \bibfield  {author} {\bibinfo {author} {\bibfnamefont {M.~J.}\ \bibnamefont
  {Stanley}}, \bibinfo {author} {\bibfnamefont {C.}~\bibnamefont {Matthiesen}},
  \bibinfo {author} {\bibfnamefont {J.}~\bibnamefont {Hansom}}, \bibinfo
  {author} {\bibfnamefont {C.}~\bibnamefont {Le~Gall}}, \bibinfo {author}
  {\bibfnamefont {C.~H.~H.}\ \bibnamefont {Schulte}}, \bibinfo {author}
  {\bibfnamefont {E.}~\bibnamefont {Clarke}}, \ and\ \bibinfo {author}
  {\bibfnamefont {M.}~\bibnamefont {Atatüre}},\ }\href {\doibase
  10.1103/physrevb.90.195305} {\bibfield  {journal} {\bibinfo  {journal} {Phys.
  Rev. B}\ }\textbf {\bibinfo {volume} {90}},\ \bibinfo {pages} {195305}
  (\bibinfo {year} {2014})}\BibitemShut {NoStop}%
\bibitem [{\citenamefont {Gschrey}\ \emph {et~al.}(2013)\citenamefont
  {Gschrey}, \citenamefont {Gericke}, \citenamefont {Sch\"{u}ßler},
  \citenamefont {Schmidt}, \citenamefont {Schulze}, \citenamefont {Heindel},
  \citenamefont {Rodt}, \citenamefont {Strittmatter},\ and\ \citenamefont
  {Reitzenstein}}]{Gschrey2013}%
  \BibitemOpen
  \bibfield  {author} {\bibinfo {author} {\bibfnamefont {M.}~\bibnamefont
  {Gschrey}}, \bibinfo {author} {\bibfnamefont {F.}~\bibnamefont {Gericke}},
  \bibinfo {author} {\bibfnamefont {A.}~\bibnamefont {Sch\"{u}ßler}}, \bibinfo
  {author} {\bibfnamefont {R.}~\bibnamefont {Schmidt}}, \bibinfo {author}
  {\bibfnamefont {J.-H.}\ \bibnamefont {Schulze}}, \bibinfo {author}
  {\bibfnamefont {T.}~\bibnamefont {Heindel}}, \bibinfo {author} {\bibfnamefont
  {S.}~\bibnamefont {Rodt}}, \bibinfo {author} {\bibfnamefont {A.}~\bibnamefont
  {Strittmatter}}, \ and\ \bibinfo {author} {\bibfnamefont {S.}~\bibnamefont
  {Reitzenstein}},\ }\href {\doibase 10.1063/1.4812343} {\bibfield  {journal}
  {\bibinfo  {journal} {Appl. Phys. Lett.}\ }\textbf {\bibinfo {volume}
  {102}},\ \bibinfo {eid} {251113} (\bibinfo {year} {2013})}\BibitemShut
  {NoStop}%
\bibitem [{\citenamefont {Gschrey}\ \emph {et~al.}(2015)\citenamefont
  {Gschrey}, \citenamefont {Thoma}, \citenamefont {Schnauber}, \citenamefont
  {Seifried}, \citenamefont {Schmidt}, \citenamefont {Wohlfeil}, \citenamefont
  {Kr\"{u}ger}, \citenamefont {Schulze}, \citenamefont {Heindel}, \citenamefont
  {Burger}, \citenamefont {Schmidt}, \citenamefont {Strittmatter},
  \citenamefont {Rodt},\ and\ \citenamefont {Reitzenstein}}]{Gschrey2015a}%
  \BibitemOpen
  \bibfield  {author} {\bibinfo {author} {\bibfnamefont {M.}~\bibnamefont
  {Gschrey}}, \bibinfo {author} {\bibfnamefont {A.}~\bibnamefont {Thoma}},
  \bibinfo {author} {\bibfnamefont {P.}~\bibnamefont {Schnauber}}, \bibinfo
  {author} {\bibfnamefont {M.}~\bibnamefont {Seifried}}, \bibinfo {author}
  {\bibfnamefont {R.}~\bibnamefont {Schmidt}}, \bibinfo {author} {\bibfnamefont
  {B.}~\bibnamefont {Wohlfeil}}, \bibinfo {author} {\bibfnamefont
  {L.}~\bibnamefont {Kr\"{u}ger}}, \bibinfo {author} {\bibfnamefont {J.~H.}\
  \bibnamefont {Schulze}}, \bibinfo {author} {\bibfnamefont {T.}~\bibnamefont
  {Heindel}}, \bibinfo {author} {\bibfnamefont {S.}~\bibnamefont {Burger}},
  \bibinfo {author} {\bibfnamefont {F.}~\bibnamefont {Schmidt}}, \bibinfo
  {author} {\bibfnamefont {A.}~\bibnamefont {Strittmatter}}, \bibinfo {author}
  {\bibfnamefont {S.}~\bibnamefont {Rodt}}, \ and\ \bibinfo {author}
  {\bibfnamefont {S.}~\bibnamefont {Reitzenstein}},\ }\href {\doibase
  10.1038/ncomms8662} {\bibfield  {journal} {\bibinfo  {journal} {Nat.
  Commun.}\ }\textbf {\bibinfo {volume} {6}},\ \bibinfo {pages} {7662}
  (\bibinfo {year} {2015})}\BibitemShut {NoStop}%
\bibitem [{\citenamefont {Schlehahn}\ \emph
  {et~al.}(2015{\natexlab{a}})\citenamefont {Schlehahn}, \citenamefont
  {Krüger}, \citenamefont {Gschrey}, \citenamefont {Schulze}, \citenamefont
  {Rodt}, \citenamefont {Strittmatter}, \citenamefont {Heindel},\ and\
  \citenamefont {Reitzenstein}}]{Schlehahn2015}%
  \BibitemOpen
  \bibfield  {author} {\bibinfo {author} {\bibfnamefont {A.}~\bibnamefont
  {Schlehahn}}, \bibinfo {author} {\bibfnamefont {L.}~\bibnamefont {Krüger}},
  \bibinfo {author} {\bibfnamefont {M.}~\bibnamefont {Gschrey}}, \bibinfo
  {author} {\bibfnamefont {J.-H.}\ \bibnamefont {Schulze}}, \bibinfo {author}
  {\bibfnamefont {S.}~\bibnamefont {Rodt}}, \bibinfo {author} {\bibfnamefont
  {A.}~\bibnamefont {Strittmatter}}, \bibinfo {author} {\bibfnamefont
  {T.}~\bibnamefont {Heindel}}, \ and\ \bibinfo {author} {\bibfnamefont
  {S.}~\bibnamefont {Reitzenstein}},\ }\href {\doibase
  http://dx.doi.org/10.1063/1.4906548} {\bibfield  {journal} {\bibinfo
  {journal} {Rev. Sci. Instrum.}\ }\textbf {\bibinfo {volume} {86}},\ \bibinfo
  {eid} {013113} (\bibinfo {year} {2015}{\natexlab{a}})}\BibitemShut {NoStop}%
\bibitem [{\citenamefont {Bylander}\ \emph {et~al.}(2003)\citenamefont
  {Bylander}, \citenamefont {Robert-Philip},\ and\ \citenamefont
  {Abram}}]{Bylander2003}%
  \BibitemOpen
  \bibfield  {author} {\bibinfo {author} {\bibfnamefont {J.}~\bibnamefont
  {Bylander}}, \bibinfo {author} {\bibfnamefont {I.}~\bibnamefont
  {Robert-Philip}}, \ and\ \bibinfo {author} {\bibfnamefont {I.}~\bibnamefont
  {Abram}},\ }\href {\doibase 10.1140/epjd/e2002-00236-6} {\bibfield  {journal}
  {\bibinfo  {journal} {EPJ D}\ }\textbf {\bibinfo {volume} {22}},\ \bibinfo
  {pages} {295} (\bibinfo {year} {2003})}\BibitemShut {NoStop}%
\bibitem [{\citenamefont {Empedocles}\ and\ \citenamefont
  {Bawendi}(1997)}]{Empedocles1997}%
  \BibitemOpen
  \bibfield  {author} {\bibinfo {author} {\bibfnamefont {S.~A.}\ \bibnamefont
  {Empedocles}}\ and\ \bibinfo {author} {\bibfnamefont {M.~G.}\ \bibnamefont
  {Bawendi}},\ }\href@noop {} {\bibfield  {journal} {\bibinfo  {journal}
  {Science}\ }\textbf {\bibinfo {volume} {278}},\ \bibinfo {pages} {2114}
  (\bibinfo {year} {1997})}\BibitemShut {NoStop}%
\bibitem [{\citenamefont {Robinson}\ and\ \citenamefont
  {Goldberg}(2000)}]{Robinson2000}%
  \BibitemOpen
  \bibfield  {author} {\bibinfo {author} {\bibfnamefont {H.~D.}\ \bibnamefont
  {Robinson}}\ and\ \bibinfo {author} {\bibfnamefont {B.~B.}\ \bibnamefont
  {Goldberg}},\ }\href {\doibase 10.1103/physrevb.61.r5086} {\bibfield
  {journal} {\bibinfo  {journal} {Phys. Rev. B}\ }\textbf {\bibinfo {volume}
  {61}},\ \bibinfo {pages} {R5086} (\bibinfo {year} {2000})}\BibitemShut
  {NoStop}%
\bibitem [{\citenamefont {Türck}\ \emph {et~al.}(2000)\citenamefont {Türck},
  \citenamefont {Rodt}, \citenamefont {Stier}, \citenamefont {Heitz},
  \citenamefont {Engelhardt}, \citenamefont {Pohl}, \citenamefont {Bimberg},\
  and\ \citenamefont {Steingrüber}}]{Tuerck2000}%
  \BibitemOpen
  \bibfield  {author} {\bibinfo {author} {\bibfnamefont {V.}~\bibnamefont
  {Türck}}, \bibinfo {author} {\bibfnamefont {S.}~\bibnamefont {Rodt}},
  \bibinfo {author} {\bibfnamefont {O.}~\bibnamefont {Stier}}, \bibinfo
  {author} {\bibfnamefont {R.}~\bibnamefont {Heitz}}, \bibinfo {author}
  {\bibfnamefont {R.}~\bibnamefont {Engelhardt}}, \bibinfo {author}
  {\bibfnamefont {U.~W.}\ \bibnamefont {Pohl}}, \bibinfo {author}
  {\bibfnamefont {D.}~\bibnamefont {Bimberg}}, \ and\ \bibinfo {author}
  {\bibfnamefont {R.}~\bibnamefont {Steingrüber}},\ }\href {\doibase
  10.1103/physrevb.61.9944} {\bibfield  {journal} {\bibinfo  {journal} {Phys.
  Rev. B}\ }\textbf {\bibinfo {volume} {61}},\ \bibinfo {pages} {9944}
  (\bibinfo {year} {2000})}\BibitemShut {NoStop}%
\bibitem [{\citenamefont {Sallen}\ \emph {et~al.}(2010)\citenamefont {Sallen},
  \citenamefont {Tribu}, \citenamefont {Aichele}, \citenamefont {André},
  \citenamefont {Besombes}, \citenamefont {Bougerol}, \citenamefont {Richard},
  \citenamefont {Tatarenko}, \citenamefont {Kheng},\ and\ \citenamefont
  {Poizat}}]{Sallen2010}%
  \BibitemOpen
  \bibfield  {author} {\bibinfo {author} {\bibfnamefont {G.}~\bibnamefont
  {Sallen}}, \bibinfo {author} {\bibfnamefont {A.}~\bibnamefont {Tribu}},
  \bibinfo {author} {\bibfnamefont {T.}~\bibnamefont {Aichele}}, \bibinfo
  {author} {\bibfnamefont {R.}~\bibnamefont {André}}, \bibinfo {author}
  {\bibfnamefont {L.}~\bibnamefont {Besombes}}, \bibinfo {author}
  {\bibfnamefont {C.}~\bibnamefont {Bougerol}}, \bibinfo {author}
  {\bibfnamefont {M.}~\bibnamefont {Richard}}, \bibinfo {author} {\bibfnamefont
  {S.}~\bibnamefont {Tatarenko}}, \bibinfo {author} {\bibfnamefont
  {K.}~\bibnamefont {Kheng}}, \ and\ \bibinfo {author} {\bibfnamefont {J.-P.}\
  \bibnamefont {Poizat}},\ }\href {\doibase 10.1038/nphoton.2010.174}
  {\bibfield  {journal} {\bibinfo  {journal} {Nature Photon.}\ }\textbf
  {\bibinfo {volume} {4}},\ \bibinfo {pages} {696} (\bibinfo {year}
  {2010})}\BibitemShut {NoStop}%
\bibitem [{\citenamefont {Houel}\ \emph {et~al.}(2012)\citenamefont {Houel},
  \citenamefont {Kuhlmann}, \citenamefont {Greuter}, \citenamefont {Xue},
  \citenamefont {Poggio}, \citenamefont {Gerardot}, \citenamefont {Dalgarno},
  \citenamefont {Badolato}, \citenamefont {Petroff}, \citenamefont {Ludwig},
  \citenamefont {Reuter}, \citenamefont {Wieck},\ and\ \citenamefont
  {Warburton}}]{Houel2012}%
  \BibitemOpen
  \bibfield  {author} {\bibinfo {author} {\bibfnamefont {J.}~\bibnamefont
  {Houel}}, \bibinfo {author} {\bibfnamefont {A.~V.}\ \bibnamefont {Kuhlmann}},
  \bibinfo {author} {\bibfnamefont {L.}~\bibnamefont {Greuter}}, \bibinfo
  {author} {\bibfnamefont {F.}~\bibnamefont {Xue}}, \bibinfo {author}
  {\bibfnamefont {M.}~\bibnamefont {Poggio}}, \bibinfo {author} {\bibfnamefont
  {B.~D.}\ \bibnamefont {Gerardot}}, \bibinfo {author} {\bibfnamefont {P.~A.}\
  \bibnamefont {Dalgarno}}, \bibinfo {author} {\bibfnamefont {A.}~\bibnamefont
  {Badolato}}, \bibinfo {author} {\bibfnamefont {P.~M.}\ \bibnamefont
  {Petroff}}, \bibinfo {author} {\bibfnamefont {A.}~\bibnamefont {Ludwig}},
  \bibinfo {author} {\bibfnamefont {D.}~\bibnamefont {Reuter}}, \bibinfo
  {author} {\bibfnamefont {A.~D.}\ \bibnamefont {Wieck}}, \ and\ \bibinfo
  {author} {\bibfnamefont {R.~J.}\ \bibnamefont {Warburton}},\ }\href {\doibase
  10.1103/PhysRevLett.108.107401} {\bibfield  {journal} {\bibinfo  {journal}
  {Phys. Rev. Lett.}\ }\textbf {\bibinfo {volume} {108}},\ \bibinfo {pages}
  {107401} (\bibinfo {year} {2012})}\BibitemShut {NoStop}%
\bibitem [{\citenamefont {Müller}\ \emph {et~al.}(2014)\citenamefont {Müller},
  \citenamefont {Bounouar}, \citenamefont {Jöns}, \citenamefont {Gläßl},\ and\
  \citenamefont {Michler}}]{Mueller2014}%
  \BibitemOpen
  \bibfield  {author} {\bibinfo {author} {\bibfnamefont {M.}~\bibnamefont
  {Müller}}, \bibinfo {author} {\bibfnamefont {S.}~\bibnamefont {Bounouar}},
  \bibinfo {author} {\bibfnamefont {K.~D.}\ \bibnamefont {Jöns}}, \bibinfo
  {author} {\bibfnamefont {M.}~\bibnamefont {Gläßl}}, \ and\ \bibinfo {author}
  {\bibfnamefont {P.}~\bibnamefont {Michler}},\ }\href {\doibase
  10.1038/nphoton.2013.377} {\bibfield  {journal} {\bibinfo  {journal} {Nature
  Photon.}\ }\textbf {\bibinfo {volume} {8}},\ \bibinfo {pages} {224} (\bibinfo
  {year} {2014})}\BibitemShut {NoStop}%
\bibitem [{\citenamefont {Galperin}\ \emph {et~al.}(2006)\citenamefont
  {Galperin}, \citenamefont {Altshuler}, \citenamefont {Bergli},\ and\
  \citenamefont {Shantsev}}]{Galperin2006}%
  \BibitemOpen
  \bibfield  {author} {\bibinfo {author} {\bibfnamefont {Y.~M.}\ \bibnamefont
  {Galperin}}, \bibinfo {author} {\bibfnamefont {B.~L.}\ \bibnamefont
  {Altshuler}}, \bibinfo {author} {\bibfnamefont {J.}~\bibnamefont {Bergli}}, \
  and\ \bibinfo {author} {\bibfnamefont {D.~V.}\ \bibnamefont {Shantsev}},\
  }\href {\doibase 10.1103/physrevlett.96.097009} {\bibfield  {journal}
  {\bibinfo  {journal} {Phys. Rev. Lett.}\ }\textbf {\bibinfo {volume} {96}},\
  \bibinfo {pages} {097009} (\bibinfo {year} {2006})}\BibitemShut {NoStop}%
\bibitem [{\citenamefont {Gardiner}\ and\ \citenamefont
  {Zoller}(2004)}]{Gardiner2004}%
  \BibitemOpen
  \bibfield  {author} {\bibinfo {author} {\bibfnamefont {C.}~\bibnamefont
  {Gardiner}}\ and\ \bibinfo {author} {\bibfnamefont {P.}~\bibnamefont
  {Zoller}},\ }\href@noop {} {\emph {\bibinfo {title} {Quantum Noise: A
  Handbook of Markovian and Non-Markovian Quantum Stochastic Methods with
  Applications to Quantum Optics}}},\ Springer Series in Synergetics\ (\bibinfo
   {publisher} {Springer},\ \bibinfo {year} {2004})\BibitemShut {NoStop}%
\bibitem [{\citenamefont {Laikhtman}(1985)}]{Laikhtman1985}%
  \BibitemOpen
  \bibfield  {author} {\bibinfo {author} {\bibfnamefont {B.~D.}\ \bibnamefont
  {Laikhtman}},\ }\href {\doibase 10.1103/physrevb.31.490} {\bibfield
  {journal} {\bibinfo  {journal} {Phys. Rev. B}\ }\textbf {\bibinfo {volume}
  {31}},\ \bibinfo {pages} {490} (\bibinfo {year} {1985})}\BibitemShut
  {NoStop}%
\bibitem [{\citenamefont {Eberly}\ \emph {et~al.}(1984)\citenamefont {Eberly},
  \citenamefont {Wódkiewicz},\ and\ \citenamefont {Shore}}]{Eberly1984}%
  \BibitemOpen
  \bibfield  {author} {\bibinfo {author} {\bibfnamefont {J.~H.}\ \bibnamefont
  {Eberly}}, \bibinfo {author} {\bibfnamefont {K.}~\bibnamefont {Wódkiewicz}},
  \ and\ \bibinfo {author} {\bibfnamefont {B.~W.}\ \bibnamefont {Shore}},\
  }\href {\doibase 10.1103/physreva.30.2381} {\bibfield  {journal} {\bibinfo
  {journal} {Phys. Rev. A}\ }\textbf {\bibinfo {volume} {30}},\ \bibinfo
  {pages} {2381} (\bibinfo {year} {1984})}\BibitemShut {NoStop}%
\bibitem [{\citenamefont {Carmichael}(1999)}]{Carmichael1999}%
  \BibitemOpen
  \bibfield  {author} {\bibinfo {author} {\bibfnamefont {H.}~\bibnamefont
  {Carmichael}},\ }\href@noop {} {\emph {\bibinfo {title} {Statistical Methods
  in Quantum Optics 1 - Master Equation and Fokker-Planck Equations}}}\
  (\bibinfo  {publisher} {Springer},\ \bibinfo {year} {1999})\BibitemShut
  {NoStop}%
\bibitem [{\citenamefont {Patel}\ \emph {et~al.}(2010)\citenamefont {Patel},
  \citenamefont {Bennett}, \citenamefont {Farrer}, \citenamefont {Nicoll},
  \citenamefont {Ritchie},\ and\ \citenamefont {Shields}}]{Patel2010}%
  \BibitemOpen
  \bibfield  {author} {\bibinfo {author} {\bibfnamefont {R.~B.}\ \bibnamefont
  {Patel}}, \bibinfo {author} {\bibfnamefont {A.~J.}\ \bibnamefont {Bennett}},
  \bibinfo {author} {\bibfnamefont {I.}~\bibnamefont {Farrer}}, \bibinfo
  {author} {\bibfnamefont {C.~A.}\ \bibnamefont {Nicoll}}, \bibinfo {author}
  {\bibfnamefont {D.~A.}\ \bibnamefont {Ritchie}}, \ and\ \bibinfo {author}
  {\bibfnamefont {A.~J.}\ \bibnamefont {Shields}},\ }\href
  {http://dx.doi.org/10.1038/nphoton.2010.161} {\bibfield  {journal} {\bibinfo
  {journal} {Nature Photon.}\ }\textbf {\bibinfo {volume} {4}},\ \bibinfo
  {pages} {632} (\bibinfo {year} {2010})}\BibitemShut {NoStop}%
\bibitem [{\citenamefont {Flagg}\ \emph {et~al.}(2010)\citenamefont {Flagg},
  \citenamefont {Muller}, \citenamefont {Polyakov}, \citenamefont {Ling},
  \citenamefont {Migdall},\ and\ \citenamefont {Solomon}}]{Flagg2010}%
  \BibitemOpen
  \bibfield  {author} {\bibinfo {author} {\bibfnamefont {E.~B.}\ \bibnamefont
  {Flagg}}, \bibinfo {author} {\bibfnamefont {A.}~\bibnamefont {Muller}},
  \bibinfo {author} {\bibfnamefont {S.~V.}\ \bibnamefont {Polyakov}}, \bibinfo
  {author} {\bibfnamefont {A.}~\bibnamefont {Ling}}, \bibinfo {author}
  {\bibfnamefont {A.}~\bibnamefont {Migdall}}, \ and\ \bibinfo {author}
  {\bibfnamefont {G.~S.}\ \bibnamefont {Solomon}},\ }\href {\doibase
  10.1103/PhysRevLett.104.137401} {\bibfield  {journal} {\bibinfo  {journal}
  {Phys. Rev. Lett.}\ }\textbf {\bibinfo {volume} {104}},\ \bibinfo {pages}
  {137401} (\bibinfo {year} {2010})}\BibitemShut {NoStop}%
\bibitem [{\citenamefont {Bernien}\ \emph {et~al.}(2012)\citenamefont
  {Bernien}, \citenamefont {Childress}, \citenamefont {Robledo}, \citenamefont
  {Markham}, \citenamefont {Twitchen},\ and\ \citenamefont
  {Hanson}}]{Bernien2012}%
  \BibitemOpen
  \bibfield  {author} {\bibinfo {author} {\bibfnamefont {H.}~\bibnamefont
  {Bernien}}, \bibinfo {author} {\bibfnamefont {L.}~\bibnamefont {Childress}},
  \bibinfo {author} {\bibfnamefont {L.}~\bibnamefont {Robledo}}, \bibinfo
  {author} {\bibfnamefont {M.}~\bibnamefont {Markham}}, \bibinfo {author}
  {\bibfnamefont {D.}~\bibnamefont {Twitchen}}, \ and\ \bibinfo {author}
  {\bibfnamefont {R.}~\bibnamefont {Hanson}},\ }\href {\doibase
  10.1103/physrevlett.108.043604} {\bibfield  {journal} {\bibinfo  {journal}
  {Phys. Rev. Lett.}\ }\textbf {\bibinfo {volume} {108}},\ \bibinfo {pages}
  {043604} (\bibinfo {year} {2012})}\BibitemShut {NoStop}%
\bibitem [{\citenamefont {Ma}\ \emph {et~al.}(2011)\citenamefont {Ma},
  \citenamefont {Zotter}, \citenamefont {Kofler}, \citenamefont {Jennewein},\
  and\ \citenamefont {Zeilinger}}]{Ma2011}%
  \BibitemOpen
  \bibfield  {author} {\bibinfo {author} {\bibfnamefont {X.-s.}\ \bibnamefont
  {Ma}}, \bibinfo {author} {\bibfnamefont {S.}~\bibnamefont {Zotter}}, \bibinfo
  {author} {\bibfnamefont {J.}~\bibnamefont {Kofler}}, \bibinfo {author}
  {\bibfnamefont {T.}~\bibnamefont {Jennewein}}, \ and\ \bibinfo {author}
  {\bibfnamefont {A.}~\bibnamefont {Zeilinger}},\ }\href {\doibase
  10.1103/PhysRevA.83.043814} {\bibfield  {journal} {\bibinfo  {journal} {Phys.
  Rev. A}\ }\textbf {\bibinfo {volume} {83}},\ \bibinfo {pages} {043814}
  (\bibinfo {year} {2011})}\BibitemShut {NoStop}%
\bibitem [{\citenamefont {Schlehahn}\ \emph
  {et~al.}(2015{\natexlab{b}})\citenamefont {Schlehahn}, \citenamefont
  {Gaafar}, \citenamefont {Vaupel}, \citenamefont {Gschrey}, \citenamefont
  {Schnauber}, \citenamefont {Schulze}, \citenamefont {Rodt}, \citenamefont
  {Strittmatter}, \citenamefont {Stolz}, \citenamefont {Rahimi-Iman},
  \citenamefont {Heindel}, \citenamefont {Koch},\ and\ \citenamefont
  {Reitzenstein}}]{Schlehahn2015a}%
  \BibitemOpen
  \bibfield  {author} {\bibinfo {author} {\bibfnamefont {A.}~\bibnamefont
  {Schlehahn}}, \bibinfo {author} {\bibfnamefont {M.}~\bibnamefont {Gaafar}},
  \bibinfo {author} {\bibfnamefont {M.}~\bibnamefont {Vaupel}}, \bibinfo
  {author} {\bibfnamefont {M.}~\bibnamefont {Gschrey}}, \bibinfo {author}
  {\bibfnamefont {P.}~\bibnamefont {Schnauber}}, \bibinfo {author}
  {\bibfnamefont {J.-H.}\ \bibnamefont {Schulze}}, \bibinfo {author}
  {\bibfnamefont {S.}~\bibnamefont {Rodt}}, \bibinfo {author} {\bibfnamefont
  {A.}~\bibnamefont {Strittmatter}}, \bibinfo {author} {\bibfnamefont
  {W.}~\bibnamefont {Stolz}}, \bibinfo {author} {\bibfnamefont
  {A.}~\bibnamefont {Rahimi-Iman}}, \bibinfo {author} {\bibfnamefont
  {T.}~\bibnamefont {Heindel}}, \bibinfo {author} {\bibfnamefont
  {M.}~\bibnamefont {Koch}}, \ and\ \bibinfo {author} {\bibfnamefont
  {S.}~\bibnamefont {Reitzenstein}},\ }\href {\doibase
  http://dx.doi.org/10.1063/1.4927429} {\bibfield  {journal} {\bibinfo
  {journal} {Appl. Phys. Lett.}\ }\textbf {\bibinfo {volume} {107}},\ \bibinfo
  {eid} {041105} (\bibinfo {year} {2015}{\natexlab{b}})}\BibitemShut {NoStop}%
\bibitem [{\citenamefont {Borri}\ \emph {et~al.}(2001)\citenamefont {Borri},
  \citenamefont {Langbein}, \citenamefont {Schneider}, \citenamefont {Woggon},
  \citenamefont {Sellin}, \citenamefont {Ouyang},\ and\ \citenamefont
  {Bimberg}}]{Borri2001}%
  \BibitemOpen
  \bibfield  {author} {\bibinfo {author} {\bibfnamefont {P.}~\bibnamefont
  {Borri}}, \bibinfo {author} {\bibfnamefont {W.}~\bibnamefont {Langbein}},
  \bibinfo {author} {\bibfnamefont {S.}~\bibnamefont {Schneider}}, \bibinfo
  {author} {\bibfnamefont {U.}~\bibnamefont {Woggon}}, \bibinfo {author}
  {\bibfnamefont {R.~L.}\ \bibnamefont {Sellin}}, \bibinfo {author}
  {\bibfnamefont {D.}~\bibnamefont {Ouyang}}, \ and\ \bibinfo {author}
  {\bibfnamefont {D.}~\bibnamefont {Bimberg}},\ }\href {\doibase
  10.1103/PhysRevLett.87.157401} {\bibfield  {journal} {\bibinfo  {journal}
  {Phys. Rev. Lett.}\ }\textbf {\bibinfo {volume} {87}},\ \bibinfo {pages}
  {157401} (\bibinfo {year} {2001})}\BibitemShut {NoStop}%
\bibitem [{\citenamefont {Förstner}\ \emph {et~al.}(2003)\citenamefont
  {Förstner}, \citenamefont {Weber}, \citenamefont {Danckwerts},\ and\
  \citenamefont {Knorr}}]{Foerstner2003}%
  \BibitemOpen
  \bibfield  {author} {\bibinfo {author} {\bibfnamefont {J.}~\bibnamefont
  {Förstner}}, \bibinfo {author} {\bibfnamefont {C.}~\bibnamefont {Weber}},
  \bibinfo {author} {\bibfnamefont {J.}~\bibnamefont {Danckwerts}}, \ and\
  \bibinfo {author} {\bibfnamefont {A.}~\bibnamefont {Knorr}},\ }\href
  {\doibase 10.1002/pssb.200303155} {\bibfield  {journal} {\bibinfo  {journal}
  {Phys. Status Solidi (b)}\ }\textbf {\bibinfo {volume} {238}},\ \bibinfo
  {pages} {419} (\bibinfo {year} {2003})}\BibitemShut {NoStop}%
\bibitem [{\citenamefont {Muljarov}\ and\ \citenamefont
  {Zimmermann}(2004)}]{Muljarov2004}%
  \BibitemOpen
  \bibfield  {author} {\bibinfo {author} {\bibfnamefont {E.~A.}\ \bibnamefont
  {Muljarov}}\ and\ \bibinfo {author} {\bibfnamefont {R.}~\bibnamefont
  {Zimmermann}},\ }\href {\doibase 10.1103/PhysRevLett.93.237401} {\bibfield
  {journal} {\bibinfo  {journal} {Phys. Rev. Lett.}\ }\textbf {\bibinfo
  {volume} {93}},\ \bibinfo {pages} {237401} (\bibinfo {year}
  {2004})}\BibitemShut {NoStop}%
\bibitem [{\citenamefont {Kaer}\ \emph
  {et~al.}(2013{\natexlab{a}})\citenamefont {Kaer}, \citenamefont {Gregersen},\
  and\ \citenamefont {Mork}}]{Kaer2013}%
  \BibitemOpen
  \bibfield  {author} {\bibinfo {author} {\bibfnamefont {P.}~\bibnamefont
  {Kaer}}, \bibinfo {author} {\bibfnamefont {N.}~\bibnamefont {Gregersen}}, \
  and\ \bibinfo {author} {\bibfnamefont {J.}~\bibnamefont {Mork}},\ }\href
  {\doibase 10.1088/1367-2630/15/3/035027} {\bibfield  {journal} {\bibinfo
  {journal} {New J. Phys.}\ }\textbf {\bibinfo {volume} {15}},\ \bibinfo
  {pages} {035027} (\bibinfo {year} {2013}{\natexlab{a}})}\BibitemShut
  {NoStop}%
\bibitem [{\citenamefont {Kaer}\ \emph
  {et~al.}(2013{\natexlab{b}})\citenamefont {Kaer}, \citenamefont {Lodahl},
  \citenamefont {Jauho},\ and\ \citenamefont {Mork}}]{Kaer2013a}%
  \BibitemOpen
  \bibfield  {author} {\bibinfo {author} {\bibfnamefont {P.}~\bibnamefont
  {Kaer}}, \bibinfo {author} {\bibfnamefont {P.}~\bibnamefont {Lodahl}},
  \bibinfo {author} {\bibfnamefont {A.-P.}\ \bibnamefont {Jauho}}, \ and\
  \bibinfo {author} {\bibfnamefont {J.}~\bibnamefont {Mork}},\ }\href@noop {}
  {\bibfield  {journal} {\bibinfo  {journal} {Phys. Rev. B}\ }\textbf {\bibinfo
  {volume} {87}},\ \bibinfo {pages} {081308} (\bibinfo {year}
  {2013}{\natexlab{b}})}\BibitemShut {NoStop}%
\bibitem [{\citenamefont {Stock}\ \emph {et~al.}(2011)\citenamefont {Stock},
  \citenamefont {Dachner}, \citenamefont {Warming}, \citenamefont {Schliwa},
  \citenamefont {Lochmann}, \citenamefont {Hoffmann}, \citenamefont {Toropov},
  \citenamefont {Bakarov}, \citenamefont {Derebezov}, \citenamefont {Richter},
  \citenamefont {Haisler}, \citenamefont {Knorr},\ and\ \citenamefont
  {Bimberg}}]{Stock2011b}%
  \BibitemOpen
  \bibfield  {author} {\bibinfo {author} {\bibfnamefont {E.}~\bibnamefont
  {Stock}}, \bibinfo {author} {\bibfnamefont {M.-R.}\ \bibnamefont {Dachner}},
  \bibinfo {author} {\bibfnamefont {T.}~\bibnamefont {Warming}}, \bibinfo
  {author} {\bibfnamefont {A.}~\bibnamefont {Schliwa}}, \bibinfo {author}
  {\bibfnamefont {A.}~\bibnamefont {Lochmann}}, \bibinfo {author}
  {\bibfnamefont {A.}~\bibnamefont {Hoffmann}}, \bibinfo {author}
  {\bibfnamefont {A.~I.}\ \bibnamefont {Toropov}}, \bibinfo {author}
  {\bibfnamefont {A.~K.}\ \bibnamefont {Bakarov}}, \bibinfo {author}
  {\bibfnamefont {I.~A.}\ \bibnamefont {Derebezov}}, \bibinfo {author}
  {\bibfnamefont {M.}~\bibnamefont {Richter}}, \bibinfo {author} {\bibfnamefont
  {V.~A.}\ \bibnamefont {Haisler}}, \bibinfo {author} {\bibfnamefont
  {A.}~\bibnamefont {Knorr}}, \ and\ \bibinfo {author} {\bibfnamefont
  {D.}~\bibnamefont {Bimberg}},\ }\href {\doibase 10.1103/PhysRevB.83.041304}
  {\bibfield  {journal} {\bibinfo  {journal} {Phys. Rev. B}\ }\textbf {\bibinfo
  {volume} {83}},\ \bibinfo {pages} {041304} (\bibinfo {year}
  {2011})}\BibitemShut {NoStop}%
\bibitem [{\citenamefont {Vagov}\ \emph {et~al.}(2011)\citenamefont {Vagov},
  \citenamefont {Croitoru}, \citenamefont {Gl\"assl}, \citenamefont {Axt},\
  and\ \citenamefont {Kuhn}}]{Vagov2011}%
  \BibitemOpen
  \bibfield  {author} {\bibinfo {author} {\bibfnamefont {A.}~\bibnamefont
  {Vagov}}, \bibinfo {author} {\bibfnamefont {M.~D.}\ \bibnamefont {Croitoru}},
  \bibinfo {author} {\bibfnamefont {M.}~\bibnamefont {Gl\"assl}}, \bibinfo
  {author} {\bibfnamefont {V.~M.}\ \bibnamefont {Axt}}, \ and\ \bibinfo
  {author} {\bibfnamefont {T.}~\bibnamefont {Kuhn}},\ }\href {\doibase
  10.1103/PhysRevB.83.094303} {\bibfield  {journal} {\bibinfo  {journal} {Phys.
  Rev. B}\ }\textbf {\bibinfo {volume} {83}},\ \bibinfo {pages} {094303}
  (\bibinfo {year} {2011})}\BibitemShut {NoStop}%
\bibitem [{\citenamefont {Santori}\ \emph {et~al.}(2009)\citenamefont
  {Santori}, \citenamefont {Fattal}, \citenamefont {Fu}, \citenamefont
  {Barclay},\ and\ \citenamefont {Beausoleil}}]{Santori2009}%
  \BibitemOpen
  \bibfield  {author} {\bibinfo {author} {\bibfnamefont {C.}~\bibnamefont
  {Santori}}, \bibinfo {author} {\bibfnamefont {D.}~\bibnamefont {Fattal}},
  \bibinfo {author} {\bibfnamefont {K.-M.~C.}\ \bibnamefont {Fu}}, \bibinfo
  {author} {\bibfnamefont {P.~E.}\ \bibnamefont {Barclay}}, \ and\ \bibinfo
  {author} {\bibfnamefont {R.~G.}\ \bibnamefont {Beausoleil}},\ }\href
  {\doibase 10.1088/1367-2630/11/12/123009} {\bibfield  {journal} {\bibinfo
  {journal} {New J. Phys.}\ }\textbf {\bibinfo {volume} {11}},\ \bibinfo
  {pages} {123009} (\bibinfo {year} {2009})}\BibitemShut {NoStop}%
\end{thebibliography}%

\end{document}